\documentclass[journal]{IEEEtran}
\usepackage{citesort}
\usepackage{amsmath,amsthm}
\usepackage{algorithmic,algorithm}
\usepackage{array}
\usepackage{amsfonts}
\usepackage{graphicx,color,overpic,psfrag,epsfig}
\usepackage{bm}
\usepackage{footmisc}
\usepackage{xcolor}
\usepackage{subfigure}
\usepackage{multirow}
\usepackage{epstopdf}

\usepackage{amssymb}

\newcommand{\qW}{{\bf W}}
\newcommand{\qI}{{\bf I}}
\newcommand{\qw}{{\bf w}}
\newcommand{\qp}{{\bf p}}
\newcommand{\qP}{{\bf P}}
\newcommand{\qq}{{\bf q}}
\newcommand{\qh}{{\bf h}}

\newcommand{\diag}{\text {diag}}

\newtheorem{lemma}{Lemma}

\begin{document}

\title{A Deep Learning Framework for Optimization  of MISO Downlink Beamforming}

\author{Wenchao Xia,~\IEEEmembership{Student~Member,~IEEE,}\thanks{W. Xia  and J. Zhang are with the Jiangsu Key Laboratory of Wireless Communications, Nanjing University of Posts and Telecommunications, Nanjing 210003, China (e-mail: 2015010203@njupt.edu.cn, zhangjun@njupt.edu.cn).} Gan Zheng,~\IEEEmembership{Senior~Member,~IEEE}\thanks{G. Zheng and Y. Zhu are with the Wolfson School of Mechanical, Electrical and Manufacturing  Engineering,  Loughborough  University,  Leicestershire,  LE11  3TU, UK (e-mail: g.zheng@lboro.ac.uk, y.zhu4@lboro.ac.uk).}, Yongxu Zhu, Jun Zhang,~\IEEEmembership{Member,~IEEE,}  Jiangzhou Wang,~\IEEEmembership{Fellow,~IEEE}\thanks{J. Wang is with the School of Engineering and Digital Arts at the University of Kent, Kent, CT2 7NT, UK (e-mail: j.z.wang@kent.ac.uk). }, and Athina P. Petropulu,~\IEEEmembership{Fellow,~IEEE}  \thanks{A. P. Petropulu is with  the Department of Electrical \& Computer Engineering Rutgers, The State University of New Jersey, Piscataway, NJ 08854  (e-mail: athinap@rutgers.edu).} }

\maketitle

\begin{abstract}
Beamforming is an effective means to improve the quality of the received signals in multiuser multiple-input-single-output (MISO) systems. Traditionally, finding the optimal beamforming solution relies on iterative algorithms, which introduces high  computational delay and is thus not suitable for real-time implementation. In this paper, we propose a deep learning framework for the optimization of downlink beamforming. In particular, the solution is obtained based on convolutional neural networks and  exploitation of expert knowledge, such as the uplink-downlink duality and the known structure of optimal solutions. Using this framework, we construct three beamforming neural networks (BNNs) for  three typical optimization problems, i.e., the signal-to-interference-plus-noise ratio (SINR) balancing problem, the power minimization problem,  and the sum rate maximization problem. For the former two problems the BNNs adopt the supervised learning approach, while  for the sum rate maximization problem  a hybrid method of supervised and unsupervised learning is employed. Simulation results show that the BNNs can achieve near-optimal solutions to the SINR balancing and power minimization problems, and a performance close to that of the weighted minimum mean squared error algorithm for the sum rate maximization problem, while in all cases enjoy  significantly reduced computational complexity. In summary, this work  paves  the way for fast realization of  optimal beamforming in multiuser MISO systems.
\end{abstract}


\section{Introduction}
Downlink beamforming techniques have attracted much attention in the past decades for their ability to realize the performance gain of multiple antennas. Beamforming has been formulated in various ways, i.e., as a signal-to-interference-plus-noise ratio  (SINR) balancing problem (also known as interference balancing problem) under a total power constraint \cite{bjornson2014optimal,boche2002general,gerlach1996base}, as a power minimization problem under quality of service (QoS) constraints \cite{shi2016sinr,rashid1998transmit,gershman2010convex,wiesel2006linear},  or as a sum rate maximization problem under a total   power constraint \cite{bjornson2014optimal,shi2011an,christensen2008weighted,Yoo2006on}.
Existing approaches to finding the optimal beamforming solutions heavily rely on  tailor-made iterative algorithms and convex optimization, which is in turn solved by general iterative algorithms such as the interior point method. For instance, the SINR balancing problem can be solved by the iterative algorithm of \cite{schubert2004solution}.
 The power minimization problem can be reformulated as a second-order cone programming (SOCP) \cite{wiesel2006linear,gershman2010convex} or semidefinite programming (SDP) problem \cite{luo2010semidefinite,bengtesson2001optimal}, which can be solved directly by an optimization software package such as CVX \cite{cvx}. Its optimal solution can also be obtained using iterative algorithms  such as Algorithm A of \cite{rashid1998joint} and the dual algorithm of \cite{shi2016sinr,schubert2004solution}.
 However, the optimal solution to the sum rate maximization problem is usually hard to obtain because the problem is nonconvex.  Locally optimal solutions are obtained via iterative algorithms, such as the weighted minimum mean squared error (WMMSE) algorithm \cite{shi2011an,christensen2008weighted}, and asymptotically optimal solutions are obtained using the water filling algorithm  combined with  zero-forcing (ZF) beamforming \cite{Yoo2006on}.

 The main drawbacks of existing iterative algorithms are the high computational complexity and the resulting latency. As a result, the beamforming technique is unable to meet the demands of real-time applications in the fifth-generation (5G) system and beyond, such as autonomous vehicles  and mission critical communications.  Even in non-real-time applications, where the small-scale fading varies in the order of milliseconds, the latency introduced by the iterative process renders the beamforming solution outdated.  To address this challenge, researchers have proposed simple heuristic beamforming solutions which admit closed-form solutions, such as the maximum-ratio transmission beamforming, the ZF beamforming, and the regularized ZF (RZF) beamforming. These heuristic beamforming solutions are directly computed based on the channel state information (CSI) without iteration, and thus involve low computational delay. However, the reduction of delay  is achieved at the cost of performance loss. The tradeoff between delay and performance seems to restrict the potential of the beamforming techniques and  its applications in practice.

 Thanks to the  recent advances in deep learning (DL) techniques, it becomes possible to find the optimal beamforming in real time by taking into account both performance and computational delay simultaneously. This is because the DL technique trains neural networks offline and  then deploys the trained neural networks for online optimization. The computational complexity is transferred from the online optimization to the offline training, and only simple linear and nonlinear operations are needed when the trained neural network is used to find the optimal beamforming solution, thus greatly reducing the computational complexity and delay.

Benefiting from the development of specialized hardware, such as graphic processing units  and field programmable gate arrays, DL can be implemented using these hardware resources conveniently. Accordingly, DL techniques have been widely used in many applications including wireless communications. A lot of research has attempted to use DL to address  physical layer issues, including channel decoding \cite{nachmani2016learning,liang2018an}, detection \cite{fan2019cnn,samuel2019learning,farsad2017detection}, channel estimation \cite{wen2018deep,wang2019deep,ye2018power}, and resource management \cite{eisen2019learning,ahmed2019deep,sun2017learning,liang2018towards,lee2018deep,van2019power,sanguinetti2018deep,chen2017echo}.
Among these efforts, the autoencoder based on unsupervised DL, investigated in \cite{Dorner2018deep,oshea2016learning}, is an ambitious attempt  to learn  an end-to-end communications system \cite{zhao2018deep}.  DL can also facilitate resource management \cite{eisen2019learning,ahmed2019deep}, including power allocation \cite{sun2017learning,liang2018towards,lee2018deep,van2019power,sanguinetti2018deep}.  Finally, \cite{zhang2019deep,zappone2019wireless} provide  an overview on the recent advances in DL-based physical layer communications and \cite{wang2017deep} suggests potential applications of DL to the physical layer.

However, with the exception of \cite{Kerret2018robust,alkhateeb2018deep,shi2018learning,huang2019unsupervised}, there are no works focusing on beamforming design in multi-antenna  communications based on DL. A common method used in the related literature is codebook-based beam selection. For example, \cite{Kerret2018robust}  designed a decentralized robust precoding scheme based on DNN in a network MIMO configuration. However, while the projection over a finite dimensional subspace  reduces the difficulty, it also results in performance loss. \cite{alkhateeb2018deep} used a DL model to predict the beamforming matrix directly from the signals received at distributed BSs  in millimeter wave systems. The sum rate performance in \cite{alkhateeb2018deep} was restricted by the quantized codebook constraint. {Different from \cite{Kerret2018robust,alkhateeb2018deep} which predicted the beamforming matrix in the finite solution space, \cite{shi2018learning,huang2019unsupervised} directly estimated the beamforming matrix; in that case the number of variables to predict increases significantly as the numbers of transmit antennas  and users increase, leading to high training complexity of the neural networks.} Furthermore, we note that none of the aforementioned works addressed the SINR balancing problem under a total power constraint, or the  power minimization problem under SINR constraints.

Motivated by the above facts and the universal approximation theorem \cite{hornik1989multilayer,zhou2019universality}, we propose a general DL framework to achieve not only near-optimal beamforming matrix, but also  reduce complexity and latency as compared to the iterative methods.  Based on the proposed framework, we develop beamforming neural networks (BNNs) to solve the three aforementioned optimization problems. Learning the optimal beamforming solution is highly nontrivial, and there are still challenges that need to be overcome in designing the BNNs. Firstly, the popular neural network software packages such as Keras and Tensorflow currently (March 2019) do not support complex numbers as input or output \cite{zhao2018deep}. However, both channel and beamforming vectors are inherently complex. {Naively using a black-box DL model to predict beamforming vectors based on CSI matrices (with a suitable real-valued representation)  will not only lead to high complexity of prediction, but also lose the specific structures of the problems of interest.}  Secondly, the power minimization problem has strict QoS constraints and guaranteeing a feasible solution using neural networks is a challenge. In addition, different from the SINR balancing and power minimization problems, there is no practically useful algorithm that can achieve the optimal solution to  the sum rate maximization problem (and other nonconvex beamforming problems), and thus the supervised learning method based on locally optimal solution cannot achieve good performance. In this paper, we  will tackle these challenges, and our main contributions are summarized as follows:
\begin{itemize}
  \item We provide a DL-based framework  for the beamforming optimization in the multiple-input-single-output (MISO)   downlink, where the BS has multiple antennas while each user terminal has a single antenna.  The proposed framework is designed based on the CNN structure.  Different from existing works where the CNN was applied to power control \cite{lee2018deep,van2019power}, resource allocation  \cite{Leem2018deep}, and wireless scheduling \cite{cui2019spatial}, the proposed framework combines a signal processing module with the neural network module by exploiting expert knowledge such as the uplink-downlink duality and  the known structure of the optimal solutions, so as to improve learning efficiency by specifying the best parameters to be learned; those parameters are typically not the direct beamforming matrix. This framework can deal with three types of beamforming optimization problems: 1) problems whose optimal solutions are easy to find and the constraints are easy to meet; 2) problems whose optimal solutions are easy to find but the constraints are hard to meet; and 3) problems which have no practically useful algorithm that can achieve optimal solutions efficiently.
      Under this framework, we propose three BNNs for solving three typical optimization problems in MISO systems, i.e., the SINR balancing problem under a total power constraint, the power minimization problem under QoS constraints, and the sum rate maximization problem under a total power constraint.
  \item In the proposed supervised BNNs for the SINR balancing and power minimization problems, instead of estimating the beamforming matrix with $NK$ elements, where $N$ is the number of the transmit antennas at the BS  and $K$ is the number of users, we exploit the uplink-downlink duality of solutions \cite{rashid1998transmit,schubert2004solution,shi2016sinr} and predict the virtual uplink power allocation vector  with only $K$ elements. Thus, the demand on the prediction capability of the BNNs in terms of  network neurons and layers  is significantly reduced. Also, the training and prediction complexity and cost are reduced. In the proposed BNN for the sum rate maximization problem, we exploit the known structure of the optimal solutions and predict two power allocation vectors with a total of $2K$ elements. This approach still has advantages as compared to predicting the beamforming matrix directly.
  \item  We  propose a hybrid two-stage BNN with both supervised and unsupervised learning to find the beamforming  solution to the sum rate maximization problem \cite{lee2018deep}, since no practically useful algorithm can find the global optimum.  In the first stage, we use the supervised learning method with a mean squared error (MSE)-based loss function to make the predictions as close as possible  to the WMMSE algorithm, which is known to achieve the  locally optimal solution. In the second stage, we modify the metric in the loss function  to be the sum rate, and update the network parameters according to the unsupervised learning method, which achieves a performance close to that of the WMMSE algorithm.
\end{itemize}

The remainder of this paper is organized as follows. Section \ref{section system model} introduces the system model and formulates three beamforming optimization problems in the MISO downlink.  Section \ref{section bnn framework} provides the framework for the beamforming optimization and then Sections \ref{section bnn for sinr balancing problem}, \ref{section bnn for power minimization problem}  and \ref{section bnn for sr maximization problem} propose the BNNs under the framework for the SINR balancing problem, the power minimization problem,  and the sum rate maximization problem, respectively.  Numerical results are presented in Section \ref{section simulation results}. Finally, conclusion is drawn in Section \ref{section conclusion}.

\textbf{Notations:} The notations are given as follows. Matrices and vectors are denoted by bold capital and lowercase symbols, respectively. $(\mathbf{A})^T$ and $(\mathbf{A})^H$  stand for transpose and conjugate transpose of $\mathbf{A}$, respectively.  The notations $||\bullet||_1$ and $||\bullet||_2$ are $l_1$ and $l_2$ norm operators, respectively. The operator $\text{diag}(\mathbf{a})$ denotes the operation to diagonalize the vector $\mathbf{a}$ into a matrix whose main diagonal elements are from $\mathbf{a}$.   Finally, $\mathbf{a}\sim\mathcal{CN}(\mathbf{0},\bm{\Sigma})$ represents a complex Gaussian vector with zero-mean and covariance matrix $\bm{\Sigma}$.

\section{System Model}\label{section system model}
We consider a downlink transmission scenario where a BS equipped with $N$ antennas serves $K$ single-antenna users. The channel between user $k$ and the BS is denoted as $\qh_k\in\mathbb{C}^{N\times 1}$. The received signal at user $k$ is given by
\begin{equation}
  y_k=\mathbf{h}^H_k \sum_{k^{\prime}=1}^K \mathbf{w}_{k^{\prime}}x_{k^{\prime}}+n_k,
\end{equation}
where $\mathbf{w}_k$  represents the beamforming vector for user $k$, $x_k\sim\mathcal{CN}(0,1)$ is the transmitted symbol from the BS to user $k$, and $n_k\sim\mathcal{CN}(0,\sigma^2)$ denotes the additive Gaussian white noise (AWGN) with zero mean and variance $\sigma^2$.
The received SINR of user $k$ equals
\begin{equation}
  \gamma^{dl}_k=\frac{|\mathbf{h}^H_k\mathbf{w}_k|^2}{\sum^K_{k^{\prime}=1,k^{\prime}\neq k}|\mathbf{h}^H_k\mathbf{w}_{k^{\prime}}|^2+\sigma^2}.
\end{equation}

One conventional optimization problem seeks to maximize $\text{min}_k \gamma^{dl}_k/\rho_k$ subject to a transmit power constraint, where $\rho_k$'s are constant weights denoting the importance of the sub-streams. Such an optimization problem is referred to as interference or SINR balancing,  and has been investigated in many works \cite{bjornson2014optimal,boche2002general,gerlach1996base}. The SINR balancing problem is formulated as:
\begin{equation}\label{p0}
\textbf{P1:}\ \max_{\qW}\min_{1\leq k\leq K}\ \ \frac{\gamma^{dl}_k}{\rho_k}, \ \
\text{s.t.}\ \sum^K_{k=1}||\mathbf{w}_k||^2\leq P_{max},
\end{equation}
where $\mathbf{W}=[\mathbf{w}_1,\mathbf{w}_2,\ldots,\mathbf{w}_K]$ is  a set of beamforming vectors and $P_{max}$ is the power budget.

Another important problem is the power minimization problem under a set of SINR constraints \cite{rashid1998transmit,gershman2010convex}. A network operator may be more interested in how to minimize the transmit power while fulfilling the demands for QoS, i.e.,
\begin{equation}\label{p1}
\textbf{P2:}\ \min_{\mathbf{W}}\ \  \sum^K_{k=1}||\mathbf{w}_k||^2, \ \
\text{s.t.}\ \gamma^{dl}_k\geq\Gamma_k,\forall k,\\
\end{equation}
where $\Gamma_k$ is the SINR constraint of user $k$. For  ease of reference, we define $\bm{\Gamma}=[\Gamma_1,\cdots,\Gamma_K]^T$ as the SINR constraint vector.

Finally, the weighted sum rate maximization problem under a total power constraint  has also attracted a lot of attention \cite{bjornson2014optimal,christensen2008weighted,shi2011an}. It can be formulated as:
\begin{equation}\label{p3}
\setlength{\abovedisplayskip}{3pt}
\setlength{\belowdisplayskip}{3pt}
\textbf{P3:}\ \max_{\mathbf{W}}\ \  \sum^K_{k=1}\alpha_k\log_2(1+\gamma^{dl}_k), \ \
\text{s.t.}\  \sum^K_{k=1}||\mathbf{w}_k||^2\leq P_{max},
\end{equation}
where $\alpha_k$ is a constant weight of user $k$.

We choose the above problems as representative examples to demonstrate the effectiveness of our proposed DL beamforming framework. Practical algorithms to find optimal solutions are available for \textbf{P1} \cite{schubert2004solution,yu2007transmitter,wiesel2006linear}  and \textbf{P2} \cite{shi2016sinr,wiesel2006linear,gershman2010convex,schubert2004solution,luo2010semidefinite}, thus supervised learning can be adopted for those problems. In this work, for simplicity, we assume that the optimal solution to problem \textbf{P2} always exists and do not consider the infeasibility of QoS constraints. Under this assumption, \textbf{P2} still has the additional challenge of satisfying  strict QoS constraints. \textbf{P3} is a difficult nonconvex  problem and is usually solved using the iterative WMMSE approach \cite{christensen2008weighted,shi2011an}, therefore,  supervised learning alone is insufficient for this case. In the rest of the paper, we will show how the solutions to these three types of problems can be efficiently learned by the proposed DL-based beamforming framework.

\section{A DL-based Framework for Beamforming Optimization}\label{section bnn framework}
\begin{figure}
\centering
\includegraphics[width=0.47\textwidth]{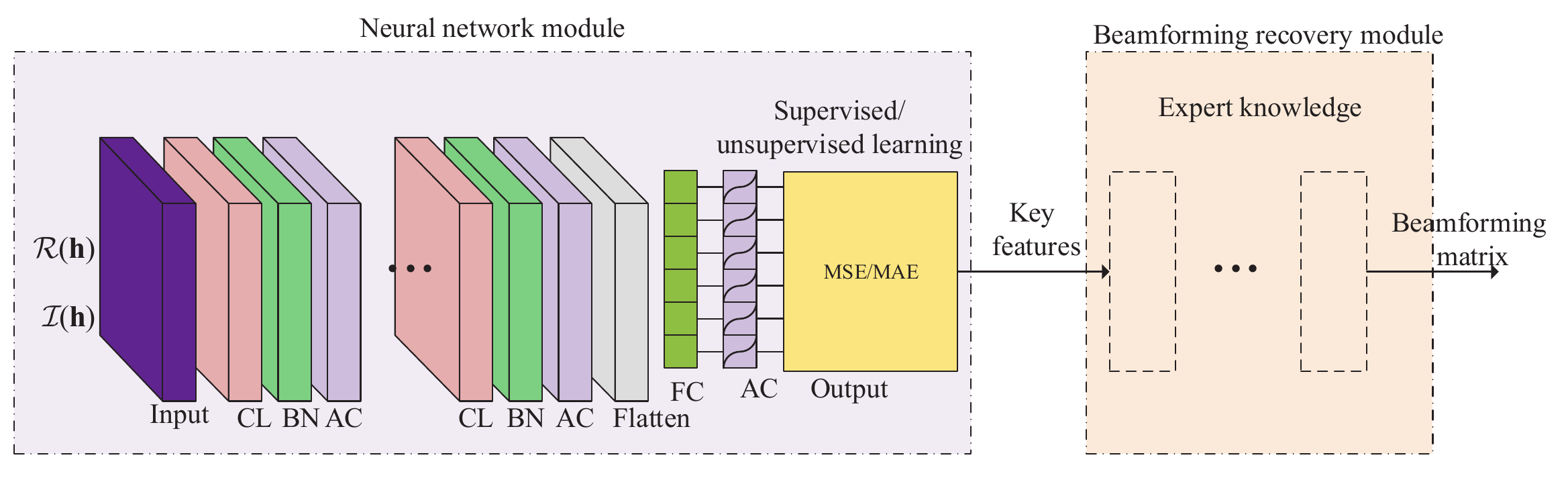}
\caption{A DL-based framework for the beamforming optimization in MISO downlink, which includes two main modules: the neural network module and the beamforming recovery module. The neural network module is composed of an input layer, convolutional (CL) layers, batch normalization (BN) layers, activation (AC) layers, a flatten layer, a fully-connected (FC) layer, and an output layer, whereas the key features and the functional layers in the beamforming recovery module are specified by the expert knowledge.}
\label{BNN framework}
\end{figure}
DL-based neural networks were initially designed for solving classification problems,  but they can also achieve satisfactory performance in regression problems. For example, the DNN was used to predict transmit power \cite{liang2018towards,sun2017learning}. Existing works mainly take real data, such as channel gains and transmit power, as input and output, but   channel  and beamforming matrices are both complex.  In addition, predicting the beamforming matrix with $NK$ elements directly may lead to inaccurate results. While we could use wider or deeper neural networks with more neurons to improve the learning ability, such huge networks would lead to high training and implementation complexity and their learning performance could not be guaranteed. For example, too deep or wide neural networks can cause over-fitting.

The proposed  DL-based framework for the beamforming optimization in MISO downlink is shown in Fig. \ref{BNN framework}. We choose the CNN architecture as the base of the framework, because the CNN has strong ability of extracting features as well as approximation ability \cite{hornik1989multilayer,zhou2019universality}. In addition,  the CNN can reduce the number of learned  parameters by sharing weights and biases \cite{van2019power}.
The proposed framework, instead of estimating the beamforming matrix directly, only predicts key features extracted from the beamforming matrix according to expert knowledge specific to the problem under consideration. Therefore, the demand for the prediction capability  in terms of network neurons and layers, as well as its complexity, is significantly reduced.

\vspace{-0.4cm}
\subsection{Structure of the Proposed Framework}
As illustrated in Fig. \ref{BNN framework},  the proposed framework is a gray-box approach that takes advantages of both the conventional signal processing and the neural network approach. The proposed framework includes two main modules: the neural network module and the beamforming recovery module. The neural network module is composed of an input layer, convolutional layers, batch normalization layers, activation layers, a flatten layer, a fully-connected  layer, and an output layer, whereas key features and the functional layers in the beamforming recovery module are specified by the expert knowledge. For ease of clarification, we assume that, besides the input, output, flatten, and fully-connected layers, there are $L=|\mathcal{L}|$ groups of functional layers in the neural network module and each group includes a convolutional layer, a batch normalization layer, and an activation layer. Below we give a brief introduction to these layers.

\subsubsection{Input Layer}
The complex channel coefficients are fed into the neural network module to predict the key features, which are not supported by the current neural network software.  To deal with this issue, two data  transformations are available. One is to separate the complex channel vector, for example $\qh=[\qh_1^T,\cdots,\qh_K^T]^T\in\mathbb{C}^{NK\times 1}$, into  in-phase component $\mathfrak{R}(\qh)$ and quadrature component $\mathfrak{I}(\qh)$, where $\mathfrak{R}(\qh)$ and $\mathfrak{I}(\qh)$ contain  the real and imaginary parts of each element in $\qh$, respectively. We call this transformation   \textbf{I/Q transformation}. Another transformation, suggested by \cite{kulin2018end}, is to map the complex channel vector $\qh$ into  two real vectors $\mathfrak{P}(\qh_k)$ and $\mathfrak{M}(\qh_k)$, where the former contains the phase information and the latter includes the magnitude information of $\qh$. This transformation is referred to as  \textbf{P/M transformation}. As far as we know, there is no evidence to show which transformation is better.  In this work, we adopt \textbf{I/Q transformation} of complex channels and formulate the input of the first convolutional layer as $[\mathfrak{R}(\qh),\mathfrak{I}(\qh)]^T\in\mathbb{R}^{2\times NK}$. Note that the samples are fed into the neural network module in batches during the training process.
\subsubsection{Convolutional Layer}
Each convolutional layer $l\in\mathcal{L}$ creates $c_l$ convolution kernels of size $a_l\times a_l$ that are convolved with the layer input $\bm{I}_{\text{conv},l}\in\mathbb{R}^{b^{(1)}_{l-1}\times b^{(2)}_{l-1}\times c_{l-1}}$, where $b^{(1)}_{l-1}$ and $b^{(2)}_{l-1}$ are the height and width of the output of the convolutional layer $l-1$, respectively. Note that $c_0=1$ $b^{(1)}_{0}=2$, and $b^{(2)}_{0}=NK$.  The parameters of the convolution kernels, including the weights $\bm{\Xi}_l\in\mathbb{R}^{a_l\times a_l\times c_l}$ and a bias vector $\bm{\xi}_l\in\mathbb{R}^{c_l\times 1}$, are shared among different elements in $\bm{I}_{\text{conv},l}$ to extract features. More specifically, the output  $\bm{O}_{\text{conv},l}\in\mathbb{R}^{b^{(1)}_{l}\times b^{(2)}_{l}\times c_{l}}$ of the convolutional layer $l$ is
\begin{equation}
\setlength{\abovedisplayskip}{3pt}
\setlength{\belowdisplayskip}{3pt}
  \bm{O}_{\text{conv},l}=\textsf{Conv}\left(\bm{I}_{\text{conv},l},\bm{\Xi}_l,\bm{\xi}_l\right), l\in\mathcal{L},
\end{equation}
where the operator $\textsf{Conv}(\cdot,\cdot,\cdot)$ denotes the convolution operation.

\subsubsection{Batch Normalization Layer}\label{BN analysis}
The batch normalization layers are introduced in the neural network module,  which can be put before or after the activation layers \cite{ioffe2015batch} according to practical experience. In the proposed framework, we adopt the former where the batch normalization layers normalize the output of the convolutional layers through subtracting the batch mean and dividing by the batch standard deviation, i.e.,
\begin{equation}
\begin{split}
  \bm{Z}_{\text{bn},l,c}[i,j]=&\frac{\bm{O}_{\text{conv},l,c}[i,j]-\mu_{l,c}}{\sqrt{\text{Var}_{l,c}+\epsilon_{l,c}}}, l\in\mathcal{L},c=1,\cdots,c_l, \\ &i=1,\cdots,b^{(1)}_{l},j=1,\cdots,b^{(2)}_{l}
\end{split}
\end{equation}
where $\bm{X}[i,j]$ denotes $(i,j)$-th element of matrix $\bm{X}$, $\bm{O}_{\text{conv},l,c}\in\mathbb{R}^{b^{(1)}_{l}\times b^{(2)}_{l}}$ is the $c$-th slice of $\bm{O}_{\text{conv},l}$, $\mu_{l,c}=\frac{\sum^F_{f=1}\sum^{b^{(1)}_{l}}_{i=1}\sum^{b^{(2)}_{l}}_{j=1}\bm{O}^{(f)}_{\text{conv},l,c}[i,j]}{Fb^{(1)}_{l}b^{(2)}_{l}}$  and $\text{Var}_{l,c}=\frac{\sum^F_{f=1}\sum^{b^{(1)}_{l}}_{i=1}\sum^{b^{(2)}_{l}}_{j=1}\big(\bm{O}^{(f)}_{\text{conv},l,c}[i,j]-\mu_{l,c}\big)^2}{Fb^{(1)}_{l}b^{(2)}_{l}}$ are the batch mean and variance of the $c$-th slice, respectively, $\epsilon_{l,c}$ is a small float added to the variance to avoid dividing by zero, and $F$ is the batch size. Note that such a simple normalization process may change what the layer can represent. To address this issue,  two trainable parameters $\theta_{l,c}$ and $\beta_{l,c}$ are introduced to scale and shift the normalized value $\bm{Z}_{\text{bn},l,c}[i,j]$  as $\hat{\bm{Z}}_{\text{bn},l,c}[i,j]=\beta_{l,c} \bm{Z}_{\text{bn},l,c}[i,j]+\theta_{l,c}$. This ``denormalization'' process is allowed by changing only these two parameters, instead of changing all parameters which may lead to the instability of the neural network module.  Besides,  the work in \cite{ioffe2015batch} claimed that the batch normalization layer can reduce the probability of over-fitting, enable a higher learning rate, and make the neural network less sensitive to the initialization of weights. Note that the batch normalization layers are element-wise functions, such that they do not change their respective input shapes.
\subsubsection{Activation Layer}
Since the predicted variables are continuous and positive real numbers, it is suggested that the activation functions that can generate negative values, such as tanh and linear functions, should not be used in the last activation layer. The rectified linear unit (ReLU) and sigmoid  functions are good choices for the last activation layer, which are given as
\begin{equation}
  \textsf{ReLU}(z)=\max(0,z) \ \text{and} \ \textsf{sigmoid}(z)=\frac{1}{1+e^{-z}},
\end{equation}
respectively. The most common choice for the intermediate activation layers is the ReLU function. Note that the functions performed in the activation layers are element-wise functions, such that their outputs have the same shapes of their inputs, respectively.
\subsubsection{Flatten Layer, Fully-connected Layer, and Output Layer}
 The flatten layer is only used to change the shape of its input into a vector, for the fully-connected layer to interpret.  The output $\mathbf{o}_{\text{fc}}\in\mathbb{R}^{m\times 1}$ of the fully-connected layer is
 \begin{equation}
\mathbf{o}_{\text{fc}}=\bm{\Pi}\bm{i}_{\text{fc}}+\bm{\pi},
 \end{equation}
where $\bm{i}_{\text{fc}}\in\mathbb{R}^{2NKc_L\times 1}$ is the input vector,  $\bm{\Pi}\in\mathbb{R}^{m\times 2NKc_L }$ and $\bm{\pi}\in\mathbb{R}^{m\times 1}$ account for the weight matrix and bias vector, respectively, and $m$ is the number of the neurons in the fully-connected layer. The main function of the output layer is to generate the predicted results after the neural network finishes training.

Note that apart from these functional layers, the loss function also plays an important role in the proposed framework,  which is marked on the output layer in Fig. \ref{BNN framework}. The loss function together with the learning rate guides the learning process of the neural network. In other words, the loss function ``tells" the neural network how to update its parameters. Since the output values are continuous, it is suggested to utilize the mean absolute error (MAE) or  the MSE as a metric. Given the predicted results of the $f$-th sample in the neural network module is $\hat{\qq}^{(f)}$ and the target result is $\qq^{(f)}$, the MAE and MSE are defined as
\begin{equation}
\text{MAE}=\frac{1}{FK}\sum_{f=1}^F{||\qq^{(f)}-\hat{\qq}^{(f)}||_1},
\end{equation}
and
\begin{equation}
\text{MSE}=\frac{1}{FK}\sum_{f=1}^F{||\qq^{(f)}-\hat{\qq}^{(f)}||^2_2},
\end{equation}
respectively. Generally speaking, the MAE function is more robust and is not affected by outliers. On the contrary,  the MSE loss function is highly sensitive to outliers in the dataset because the MSE function tries to adjust the model according to these outlier values, at the expense of other samples \cite{botchkarev2019performance}. In this work, the training dataset is  generated by simulations and  outliers are not an issue.  Then we choose the MSE as the loss metric because its gradient is easier to  calculate than that of the MAE.

\subsubsection{Beamforming Recovery Module}
The beamforming recovery module is an important component whose aim is to recover the beamforming matrix  from the predicted key features at the output layer. The functional layers in the beamforming recovery module are designed according to the expert knowledge of the beamforming optimization which maps/converts   the key features to the beamforming matrix. The expert knowledge is problem-dependent and has no unified form, but what is in common is that the expert knowledge can significantly reduce the number of variables to be predicted compared to the beamforming matrix.  For example, the uplink-downlink duality and specific solution structures  are the typical expert knowledge for beamforming optimization.

The key features should be chosen carefully to meet some constraints required by applying the universal approximation theorem \cite{hornik1989multilayer,sun2017learning}, so that a feedforward network exists which can approximate the continuous mapping from the channel coefficients to the key features. More specifically, assume that $\bm{\tau}$ is a vector containing the chosen key features, the mapping function $f(\bullet)$ from $\qh$ to $\bm{\tau}$, i.e., $\bm{\tau}=f(\qh)$, should be a real-valued continuous function over a compact set. The compact set requirement holds whenever the possible values of the input $\qh$ are bounded. However,  the continuity of the mapping function depends on the choice of the key features.

In next three sections we will propose three BNNs under the proposed framework for problems \textbf{P1}, \textbf{P2}, and \textbf{P3}, respectively, and provide implementation details to show how to make use of the expert knowledge and choose the key features.
\vspace{-0.4cm}
\subsection{Computational Complexity}
The computational complexity of the proposed framework involves two main tasks: the online prediction and  the offline training. To the best of our knowledge, complexity analysis of the offline training is still an open issue mainly because of the complex implementation of the backpropagation process. However, since the training  is performed offline, and updated at a much longer time-scale compared to the online prediction, we assume its complexity can be afforded \cite{matthiesen2018deep}. Thus, we focus on the complexity of  the online prediction. In addition,  the functional layers  are problem-dependent in the beamforming recovery module, so only the complexity of the neural network module is analyzed below.

Big-O notation is a common method to describe the complexity of an algorithm. Given there are $c_l$ kernels of size $a_l\times a_l$ in the $l$-th convolutional layer, then the numbers of multiplication and addition operations of convolutional layer $l$ are the same and equal to $a^2_l b^{(1)}_lb^{(2)}_l c_{l-1}c_l$. Thus, the total time complexity of all convolutional layers measured by the number of multiplications  is $\mathcal{O}\left(\sum_{l\in\mathcal{L}}a^2_l b^{(1)}_lb^{(2)}_l c_{l-1}c_l\right)$ \cite{he2015convolutional}. It is known that the batch normalization layers and activation layers are element-wise functions, thus the computational complexity of total batch normalization layers and total activation layers in $L$ groups is  $\mathcal{O}\left(\sum_{l\in\mathcal{L}} b^{(1)}_lb^{(2)}_l c_l\right)$.
The numbers of multiplication and addition operations of the fully-connected layer are also the same and equal to $b^{(1)}_Lb^{(2)}_L c_Lm$, respectively. Then the time complexity of the fully-connected layer is given as $\mathcal{O}\left(b^{(1)}_Lb^{(2)}_L c_Lm\right)$. Besides, the complexity of the input, output, and flatten layers are ignored due to the simplicity of their functions. If all convolutional layers use the kernels of size $3\times 3$ and apply
stride 1 and zero padding 1, then $b^{(1)}_{l}=2$ and $b^{(2)}_{l}=NK, \forall l\in \mathcal{L}$. Based on the above analysis and assuming the parameters of the neural network module are fixed, predicting the output of the neural network module needs  $2NK\sum_{l\in\mathcal{L}} (9c_lc_{l-1}+c_l)+2NKc_Lm+2m $  arithmetic operations including multiplications, divisions, and exponentiations,  and has an approximate complexity $\mathcal{O}\left(NK\right)$.
\vspace{-0.2cm}
\section{BNN for SINR Balancing Problem}\label{section bnn for sinr balancing problem}
As mentioned above, estimating the beamforming matrix directly leads to the higher complexity of prediction due to the large amount of variables. In order to reduce the prediction complexity, we introduce a scheme which first predicts the power allocation vector as the key feature and then achieves the corresponding beamforming matrix based on the predicted results. Such a scheme is based on the expert knowledge named the uplink-downlink duality.
\vspace{-0.4cm}
\subsection{Uplink-Downlink Duality}
Before we present  the BNN for the SINR balancing problem \textbf{P1}, we first introduce the following  lemma to describe the uplink-downlink duality of problem \textbf{P1} \cite{schubert2004solution}.
\begin{lemma}
Given $\tilde{\qW}=[\tilde{\qw}_1,\tilde{\qw}_2,\ldots,\tilde{\qw}_K]$ and $P_{max}$, we have
\begin{equation}
  C^{dl}(\tilde{\mathbf{W}},P_{max})= C^{ul}(\tilde{\mathbf{W}},P_{max}),
\end{equation}
where $C^{dl}(\tilde{\qW},P_{max})$ and $C^{ul}(\tilde{\qW},P_{max})$ are given as
\begin{align}\label{equivalent problem of p1}
  C^{dl}(\tilde{\qW},P_{max})=&\max_{\qp}\min_{1\leq k\leq K}\frac{\gamma^{dl}_k(\tilde{\qW},\qp)}{\rho_k}\\
  \text{s.t.}\ &||\qp||_1\leq P_{max},\nonumber\\
  &||\tilde{\qw}_k||_2=1, \forall k,\nonumber
\end{align}
and
\begin{align}\label{ul_sinr_pro}
  C^{ul}(\tilde{\qW},P_{max})=&\max_{\qq}\min_{1\leq k\leq K}\frac{\gamma^{ul}_k(\tilde{\qW},\qq)}{\rho_k}\\
  \text{s.t.}\ &||\mathbf{q}||_1\leq P_{max},\nonumber\\
  & ||\tilde{\qw}_k||_2=1, \forall k,\nonumber
\end{align}
respectively, with
\begin{equation}
  \gamma^{dl}_k(\tilde{\qW},\qp)=\frac{p_k|\qh^H_k\tilde{\qw}_k|^2}{\sum^K_{k^{\prime}=1,k^{\prime}\neq k}p_{k^{\prime}}|\qh^H_k\tilde{\qw}_{k^{\prime}}|^2+\sigma^2},
\end{equation}
and
\begin{equation}\label{uplink sinr}
  \gamma^{ul}_k(\tilde{\qW},\qq)=\frac{q_k|\qh^H_k\tilde{\qw}_k|^2}{\sum^K_{k^{\prime}=1,k^{\prime}\neq k}q_{k^{\prime}}|\qh^H_{k{\prime}}\tilde{\qw}_{k}|^2+\sigma^2}.
\end{equation}
Note that $\qp=[p_1,\ldots,p_K]^T$  and $\mathbf{q}=[q_1,\ldots,q_K]^T$ are downlink and uplink power vectors, respectively\footnote{Lemma 1 can be easily extended to the case with non-identical noise power levels. More details can refer to \cite{schubert2004solution}.}.
\end{lemma}
Note that problem \eqref{equivalent problem of p1} is an equivalent virtual problem of problem \textbf{P1} whose optimal solutions are connected by $\qW^{\ast}=\tilde{\qW}^{\ast}\qP^{\ast}$ where $\qP^{\ast}=\diag(\qp^{\ast})$, {{$\qW^{\ast}$ is the optimal solution to problem \textbf{P1}, and $\tilde{\qW}^{\ast}$ and $\qp^{\ast}$ are the optimal solutions to problem \eqref{equivalent problem of p1}}}.
Based on \textbf{Lemma 1}, we find that the uplink and downlink scenarios have the same achievable SINR  region and the normalized beamforming  designed for the uplink reception immediately carries over to the downlink transmission \cite{schubert2004solution}. Thus we first obtain the optimal power allocation $\mathbf{q}^{\ast}$ and beamforming matrix $\tilde{\qW}^{\ast}$ for the easier-to-solve uplink problem \eqref{ul_sinr_pro} instead of the downlink problem \eqref{equivalent problem of p1}. Then given the optimal beamforming $\tilde{\qW}^{\ast}$, the optimal $\mathbf{p}^{\ast}$ is obtained as the first $K$ components of the dominant eigenvector of the following matrix \cite{yang1998optimal}
\begin{equation}\label{transform matrix}
\bm{\Upsilon}(\tilde{\qW}^{\ast},P_{max})=
\left[
\begin{array}{cc}
\mathbf{D}\mathbf{U} & \mathbf{D}\bm{\sigma} \\
\frac{1}{P_{max}}\mathbf{1}^T\mathbf{D}\mathbf{U} & \frac{1}{P_{max}}\mathbf{1}^T\mathbf{D}\bm{\sigma} \\
\end{array}
\right],
\end{equation}
where $\bm{\sigma}=\sigma^2\mathbf{1}$, $\mathbf{1}=[1,1,\ldots,1]^T\in\mathbb{R}^{K\times 1}$,
$\mathbf{D}=\diag\{\rho_1/|(\tilde{\qw}^{\ast}_1)^H\qh_1|^2,\ldots,\rho_K/|(\tilde{\qw}^{\ast}_K)^H\qh_K|^2\}$, and
\begin{equation}
[\mathbf{U}]_{kk^{\prime}}=
\begin{cases}
|(\tilde{\qw}^{\ast}_{k^{\prime}})^H\qh_k|^2,& \text{if}\ k^{\prime}\neq k,\\
0,&\text{else}.
\end{cases}
\end{equation}
Finally, the downlink beamforming matrix is derived as $\qW^{\ast}=\tilde{\qW}^{\ast}\qP^{\ast}$. Thus, instead of predicting $\qW$ directly, we can predict the uplink power allocation vector $\mathbf{q}$. In the supervised learning method,  the prediction performance of the BNN depends on the quality of training samples. To generate the training samples, the optimal $\qq^{\ast}$ and $\tilde{\qW}^{\ast}$ can be found by an iterative optimization algorithm in \cite[Table 1]{schubert2004solution}.

Note that $\bm{\Upsilon}(\tilde{\qW}^{\ast},P_{max})$ is a non-negative matrix and the optimal objective value of problem \textbf{P1} is the reciprocal of the largest eigenvalue of $\bm{\Upsilon}(\tilde{\qW}^{\ast},P_{max})$ \cite{yang1998optimal}. According to the Perron-Frobenius theory, for any nonnegative real matrix $\bm{\Omega}$ with spectral radius $\chi(\bm{\Omega})$, there exist a vector $\bm{\delta}\geq 0$ such that $\bm{\Omega}\bm{\delta}=\chi(\bm{\Omega})\bm{\delta}$ \cite{seneta2006non}.  Based on
\cite[Theorem 3]{schubert2004solution}, the sequence of the target value of problem \textbf{P1} provided by the iterative algorithm in \cite[Table 1]{schubert2004solution} is strictly monotonically increasing and the largest eigenvalue of $\bm{\Upsilon}(\tilde{\qW}^{\ast},P_{max})$ is unique. Then  the corresponding eigenvector containing $\qq$ is a continuous and bounded function of $\qh$ according to \cite[Chapter 3]{ortega1990numerical}. Thus, we can use a neural network to approximate the mapping function from $\qh$ to $\qq$ \cite{hornik1989multilayer}.

\subsection{BNN Structure}\label{subsection bnn structure of sinr}
\begin{figure}
\centering
\includegraphics[width=0.47\textwidth]{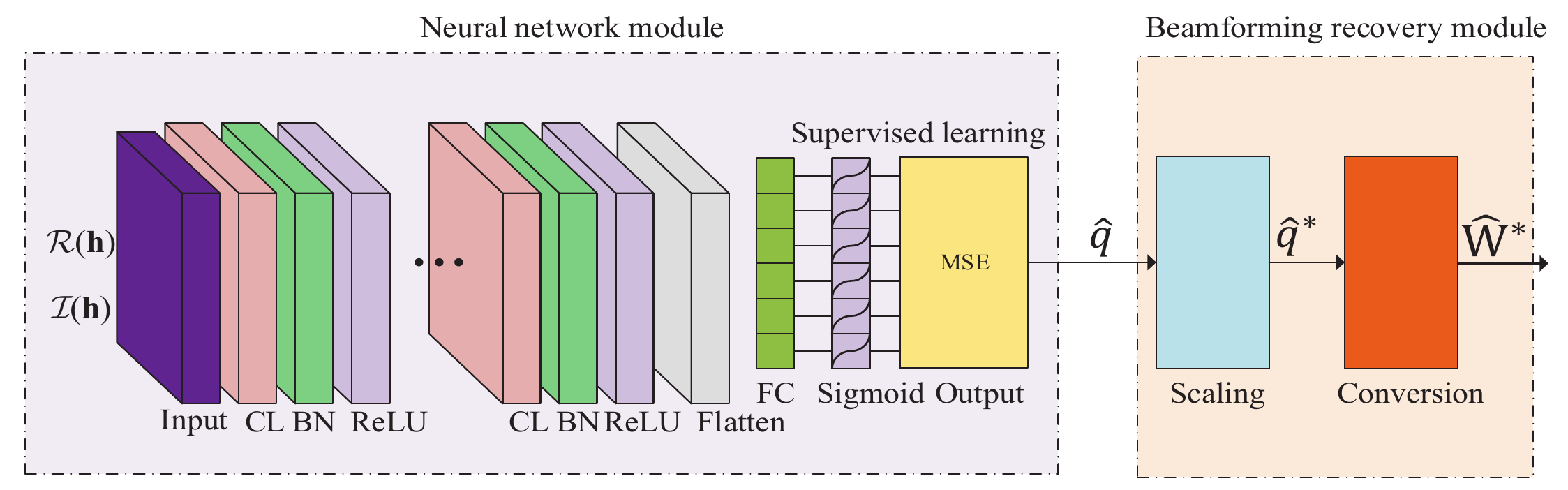}
\caption{BNN for the SINR balancing problem.}
\label{BNN structure for sinr}
\end{figure}

The proposed BNN for  problem \textbf{P1}, shown in Fig. \ref{BNN structure for sinr}, is based on the proposed BNN framework in Fig. \ref{BNN framework}. The functions and operations of the basic layers such as the input, convolutional, batch normalization, and output layers, are the same as those in the proposed framework. Therefore, we do not explain these layers here and readers can refer to Section \ref{section bnn framework} for detail. Note that in the proposed BNN for problem \textbf{P1}, the intermediate activation layers are fulfilled with the ReLU function whereas the last activation layer is implemented using the sigmoid function. Besides the existing layers in the framework, a scaling layer and a conversion layer are also introduced in the BNN for problem \textbf{P1}, which belong to the beamforming recovery module. In the following, we give the details of the scaling layer and the conversion layer.
\subsubsection{Scaling Layer}
Due to the existence of prediction error, it is almost impossible to guarantee that the output of the output layer always meets the power constraint in problem \textbf{P1}.  According to \cite{Visotsky199optimum}, the optimal solution is achieved  when the equality of the constraint in problem  \textbf{P1} holds. Therefore, we scale the results of the output layer $\hat{\mathbf{q}}$ to meet the power constraint by the following transformation,
\begin{equation}
\hat{\mathbf{q}}^{\ast}=\frac{P_{max}}{||\hat{\mathbf{q}}||_1}\hat{\mathbf{q}}.
\end{equation}
\subsubsection{Conversion Layer}
After receiving the scaled power allocation vector $\hat{\mathbf{q}}^{\ast}$, we can achieve the downlink beamforming matrix $\hat{\qW}^{\ast}$ as the final output of the BNN based on $\hat{\qq}^{\ast}$ by the conversion layer. The beamforming recovery implemented by the conversion layer includes the following process:
\begin{enumerate}
  \item Calculate $\mathbf{T}^{\ast}=\sigma^2\qI_N+\sum^K_{k=1}\hat{q}^{\ast}_k\qh_k\qh^H_k$.
  \item Calculate $\tilde{\qw}^{\ast}_k=\tilde{\qw}^{\ast}_k/||\tilde{\qw}^{\ast}_k||_2, \forall k,$ where $\tilde{\qw}^{\ast}_k=(\mathbf{T}^{\ast})^{-1}\qh_k$.
  \item  Find the maximal eigenvalue $\psi_{max}^{\ast}$ of $\bm{\Upsilon}(\tilde{\qW}^{\ast},P_{max})$ and the associated eigenvector  with respect to $\psi_{max}^{\ast}$, i.e., $\bm{\Upsilon}(\tilde{\qW}^{\ast},P_{max})\bigl[\begin{smallmatrix}\hat{\qp}^{\ast}\\ 1\end{smallmatrix}\bigr]=\psi_{max}^{(i)}\bigl[\begin{smallmatrix}\hat{\qp}^{\ast}\\1\end{smallmatrix}\bigr]$.
  \item Output $\hat{\qW}^{\ast}=\tilde{\qW}^{\ast}\hat{\qP}^{\ast}$ as the final result where $\hat{\qP}^{\ast}=\text{diag}(\hat{\qp}^{\ast})$.
\end{enumerate}

Note that the time complexity of  the beamforming recovery module is $\mathcal{O}(KN^2+N^3+K^3)$. In the proposed BNN for the SINR balancing problem \textbf{P1}, the supervised learning with the loss function based on the MSE metric is adopted.
\vspace{-0.4cm}
\section{BNN for Power Minimization Problem}\label{section bnn for power minimization problem}
Similar to the BNN for the SINR balancing problem \textbf{P1}, the BNN for the power minimization problem \textbf{P2} obtains the downlink beamforming matrix according to the uplink-downlink duality, i.e., the expert knowledge. Specifically, we first predict the uplink power allocation vector as the key features using the trained neural network, then obtain the normalized beamforming matrix based on the predicted results. Finally, the downlink beamforming matrix is recovered from the normalized beamforming matrix by the uplink-downlink conversion method.

\subsection{Uplink-Downlink Duality}
Note that the conversion method adopted in the BNN for  problem \textbf{P1} can not be used again, because the power budget $P_{max}$ is unknown in the power minimization problem \textbf{P2}.  Instead, we employ the conversion method in the following lemma \cite{yu2007transmitter}.

\begin{lemma}
Given the optimal beamforming matrix $\tilde{\qW}^{\ast}=[\tilde{\qw}^{\ast}_1,\ldots,\tilde{\qw}^{\ast}_K]$ for the uplink problem\footnote{In this work, for simplicity, we assume the solution to problem \textbf{P2} always exists. However, it can happen that the wireless network only satisfies some of the users and thus the user selection is needed. To address this issue, a possible solution is to train another neural network for user selection, and then optimize the beamforming matrix among the selected users.}, i.e.,
\begin{equation}
\begin{split}
&\min_{\mathbf{q},\tilde{\qW}}\sum_{k=1}^K q_k\\
  \text{s.t.}\ &\gamma^{ul}_k(\tilde{\qW},\mathbf{q})\geq \Gamma_k,\\
  & ||\tilde{\qw}_k||_2=1, \forall k,
\end{split}
\end{equation}
where $ \gamma^{ul}_k(\tilde{\qW},\mathbf{q})$ is given as in \eqref{uplink sinr}.

The optimal beamforming vectors $\qw^{\ast}_k, \forall k,$ for the downlink problem \textbf{P2}, can be obtained by multiplying  the optimal normalized beamforming vector $\tilde{\qw}^{\ast}_k$ by a scaling factor, i.e., $\qw^{\ast}_k=p^{\ast}_k\tilde{\qw}_k^{\ast},\forall k$, where $p^{\ast}_k$ is the $k$-th element of vector $\qp^{\ast}=[p^{\ast}_1,\ldots,p^{\ast}_K]^T\in\mathbb{R}^{K\times 1}$ and
\begin{equation}\label{optimal downlink power}
  \qp^{\ast}=\sigma^2\bm{\Psi}^{-1}\mathbf{1},
\end{equation}
where
\begin{equation}
[\bm{\Psi}]_{kk^{\prime}}=
\begin{cases}
\frac{1}{\Gamma_k}|\qh^H_k\tilde{\qw}^{\ast}_k|^2,\ \text{if} \ k=k^{\prime}, \\
-|\qh^H_k\tilde{\qw}^{\ast}_{k^{\prime}}|^2, \ \text{else}.
\end{cases}
\end{equation}
\end{lemma}
The vector $\qp^{\ast}$ of the scaling factors   is the optimal downlink power allocation vector.  Given the optimal normalized beamforming matrix $\tilde{\qW}^{\ast}$, \textbf{Lemma 2} allows us to achieve the optimal  downlink power vector $\qp^{\ast}$ by \eqref{optimal downlink power}, then  $\qW^{\ast}=\tilde{\qW}^{\ast}\qP^{\ast}$. Actually, if we know the uplink power allocation vector $\qq$, the normalized beamforming matrix $\tilde{\qW}$ can be inferred as
\begin{equation}
 \tilde{\qw}_k=\frac{\mathbf{T}^{-1}\qh_k}{||\mathbf{T}^{-1}\qh_k||_2},\forall k,
\end{equation}
where $\mathbf{T}=\sigma^2\qI_N+\sum^K_{k=1}q_k\qh_k\qh^H_k$. Therefore, the only results that need to be predicted by the BNN is the uplink power allocation vector $\qq$, which reduces significantly the computational complexity compared to the strategy that attempts to predict the beamforming matrix directly. The iterative algorithm in \cite{shi2016sinr} provides a way to achieve the optimal $\qq^{\ast}$ as the training samples in the supervised learning method. Besides, such an iterative algorithm suggests  the mapping function from $\qh$ to $\qq$ is continuous  \cite[Theorem 1]{sun2017learning}, so it can be approximated by a neural network.

\subsection{BNN Structure}

\begin{figure}
\centering
\includegraphics[width=0.47\textwidth]{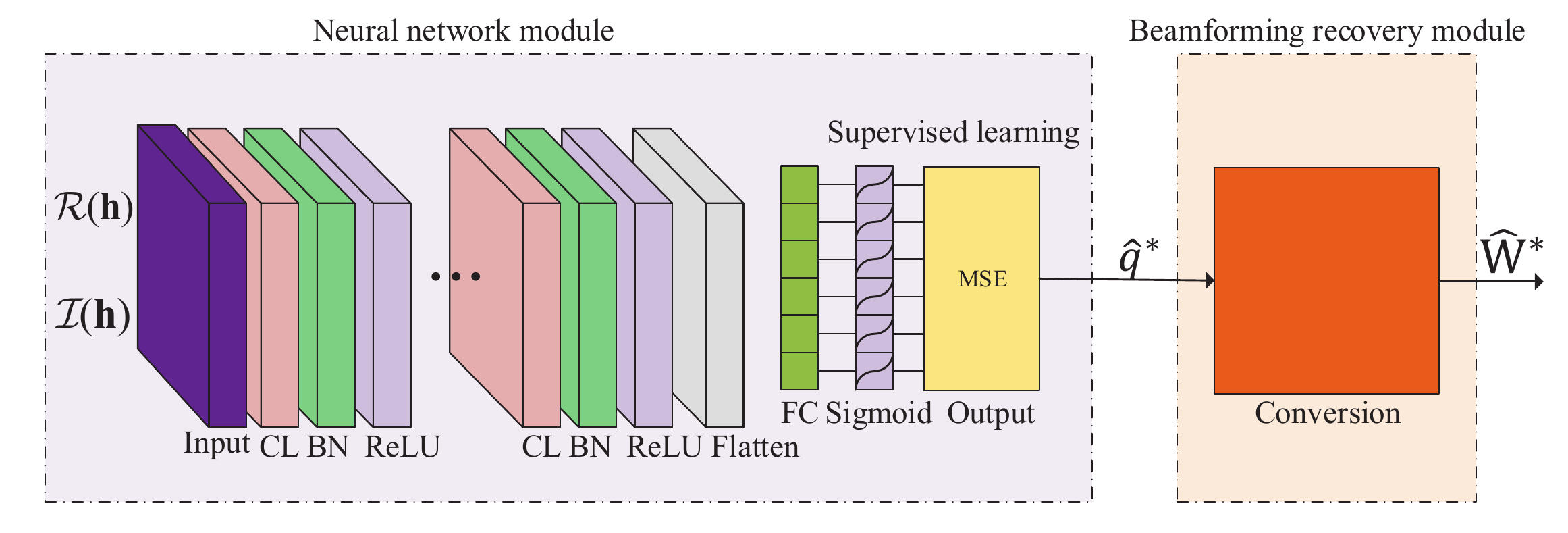}
\caption{BNN for the power minimization problem.}
\label{BNN structure for power}
\end{figure}

The BNN  for problem \textbf{P2} in Fig. \ref{BNN structure for power} is also based on the proposed BNN framework. However,   the operations of the conversion layer in Fig. \ref{BNN structure for power} are different from those in the BNN for problem \textbf{P1}. After receiving the uplink power allocation vector $\hat{\qq}^{\ast}$ from the output layer, the beamforming recovery in the conversion layer performs the following operations:
\begin{enumerate}
  \item Calculate $\mathbf{T}^{\ast}=\sigma^2\qI_N+\sum^K_{k=1}\hat{q}^{\ast}_{k}\qh_{k}\qh^H_{k}$.
  \item Calculate $\tilde{\qw}^{\ast}_k=\tilde{\qw}^{\ast}_k/||\tilde{\qw}^{\ast}_k||_2, \forall k,$ where $\tilde{\qw}^{\ast}_k=(\mathbf{T}^{\ast})^{-1}\qh_k$.
  \item Calculate the downlink power allocation vector $\hat{\qp}^{\ast}=\sigma^2(\bm{\Psi}^{\ast}(\tilde{\qW}^{\ast},\bm{\Gamma}))^{-1}\mathbf{1}$.
  \item   Output the downlink beamforming vectors $\hat{\qw}^{\ast}_k=\hat{p}^{\ast}_k\tilde{\qw}^{\ast}_k, \forall k,$ as the final results.
\end{enumerate}

Here, the time complexity of the beamforming recovery module is $\mathcal{O}(KN^2+N^3+K^3)$. Note that the predicted power vector $\hat{\qq}^{\ast}$ by the BNN is, in general,  not exact.
The prediction error will lead to the inaccuracy of power allocation vector $\hat{\qp}^{\ast}$ as well as the downlink beamforming $\hat{\qW}^{\ast}$. More specifically, if the predicted power vector $\hat{\qq}^{\ast}$  has an acceptable accuracy with respect to the target power vector $\qq^{\ast}$, i.e., $||\qq^{\ast}-\hat{\qq}^{\ast}||^2_2<\varepsilon$ where $\varepsilon$ is  a small constant, then we can obtain a suboptimal solution whose objective value is larger than that of the  optimal solution, i.e., $\sum^K_{k=1}||\hat{\qw}^{\ast}_k||^2_2>\sum^K_{k=1}||\qw^{\ast}_k||^2_2$. Intuitively, the extra power consumption $q_{extra}=\sum^K_{k=1}||\hat{\qw}^{\ast}_k||^2_2-\sum^K_{k=1}||\qw^{\ast}_k||^2_2$ can be regarded as the cost of the prediction error. However, if  the predicted vector $\hat{\qq}^{\ast}$  has a significant error, i.e., $||\qq^{\ast}-\hat{\qq}^{\ast}||^2_2\gg\varepsilon$, the downlink beamforming $\hat{\qW}^{\ast}$ inferred from the prediction $\hat{\qq}^{\ast}$ may become infeasible since some elements of the vector $\hat{\qp}^{\ast}$ have negative values. This suggests that different from problem \textbf{P1}, there is a certain  probability of  infeasibility  of the BNN prediction for problem \textbf{P2}. However, our experiments show that the failure probability of the proposed BNN for problem \textbf{P2} is lower than $1\%$ in most settings. More details will be given in Section \ref{section simulation results}. Moreover, the supervised learning with the loss function based on the MSE metric is adopted in the proposed BNN for problem \textbf{P2}.

\section{BNN for Sum Rate Maximization Problem}\label{section bnn for sr maximization problem}
Different from the SINR balancing problem \textbf{P1} and the power minimization problem \textbf{P2}, no practically useful algorithm is available to find the optimal solution to the sum rate maximization problem \textbf{P3}, for which one cannot make use of uplink-downlink duality directly.  However, we will exploit  a connection  between problems \textbf{P2} and \textbf{P3} to find some key features of the optimal solution to problem \textbf{P3}.
\vspace{-0.4cm}
\subsection{Solution Structure}
A fact was mentioned in \cite{bjornson2013optimal} that the optimal solution to problem \textbf{P2}, using the minimal amount of power to achieve the given SINR targets, must meet the power constraint in problem \textbf{P3} to achieve the maximal sum rate. More specifically, given the optimal transmit power $P^{\star}$ of problem \textbf{P2} and setting the total power constraint $P_{max}$ in problem \textbf{P3} as $P^{\star}$,  the SINR values of each user in problem \textbf{P3} can be calculated. By setting the SINR targets in problem \textbf{P2} with these calculated SINR values,  the solutions to problems \textbf{P2} and \textbf{P3} will be the same. According to the connection between problems \textbf{P2} and \textbf{P3}, it has been pointed out in \cite{bjornson2014optimal}  that the optimal downlink beamforming vectors  for problem \textbf{P3} follows the structure as
\begin{equation}\label{solution struc of sumrate}
  \qw_k^{\ast}=\sqrt{p_k}\frac{(\qI_N+\sum^K_{k=1}{\frac{\lambda_k}{\sigma^2}\qh_k\qh_k^H})^{-1}\qh_k}{||(\qI_N+\sum^K_{k=1}{\frac{\lambda_k}{\sigma^2}\qh_k\qh_k^H})^{-1}\qh_k||_2}, \forall k,
\end{equation}
where $\lambda_k$ is a positive parameter and $\sum_{k=1}^K\lambda_k=\sum_{k=1}^Kp_k=P_{max}$ according to the strong duality of problem \textbf{P2}. This is because $P_{max}$ is the optimal cost function in problem \textbf{P2} and $\sum_{k=1}^K\lambda_k$ is the dual function. Note that the parameter vector $\bm{\lambda}=[\lambda_1,\ldots,\lambda_K]^T$ can be considered as a virtual power allocation vector. The solution structure in \eqref{solution struc of sumrate} provides the required expert knowledge for the beamforming design in problem \textbf{P3} and $\bm{\lambda}$ and $\qp$ are the key features.  But to our best knowledge, there is no low-complexity algorithm in the literature that can find the optimal $p^{\ast}_k$ and $\lambda^{\ast}_k$ in \eqref{solution struc of sumrate}. {An improved and faster branch-and-bound algorithm  was developed in \cite{matthiesen2018deep,zappone2019wireless} to find the globally optimal solution,  but it is mostly effective for power control problems.} The WMMSE algorithm is a good choice to find the locally optimal solutions \cite{shi2011an,christensen2008weighted}, and such an iterative algorithm ensures the continuity of the mapping from the channel to the solution, and can be learned by a neural network \cite{sun2017learning,van2019power}. Therefore, we can obtain the power allocation vectors $\qp$ and $\bm{\lambda}$ according to the WMMSE algorithm. The supervised learning with the loss function based on the MSE metric will be first used to achieve as close to the results of the WMMSE algorithm as possible, i.e.,
\begin{equation}\label{loss1}
  \text{Loss}=\frac{1}{2LK}\sum_{l=1}^L{\left(||\underline{\qp}^{(l)}-\hat{\qp}^{(l)}||^2_2+||\underline{\bm{\lambda}}^{(l)}-\hat{\bm{\lambda}}^{(l)}||^2_2\right)},
\end{equation}
where $\underline{\qp}^{(l)}$ and $\underline{\bm{\lambda}}^{(l)}$ are the power vectors obtained from the WMMSE algorithm, and $\hat{\qp}^{(l)}$ and $\hat{\bm{\lambda}}^{(l)}$ are the predicted results of the BNN. It is worth pointing out that the results in the training samples of problems \textbf{P1} and \textbf{P2} are optimal, thus the MSE-based loss function is equivalent to the objective function and the supervised learning method updates network parameters towards the direction of the optimal solution. However, the WMMSE algorithm for problem \textbf{P3} is locally optimal and thus \eqref{loss1} is not equivalent to the real objective of problem \textbf{P3} which aims to maximize the weighted sum rate. To further improve the sum rate performance, we continue to train the BNN in an unsupervised learning way, whose loss function takes the objective function directly as a metric, i.e.,
\begin{equation}\label{loss2}
  \text{Loss}=-\frac{1}{2KL}\sum^L_{l=1}\sum^K_{k=1}\alpha^{(l)}_k\log_2\left(1+\gamma^{ul,(l)}_k\right).
\end{equation}

\subsection{Hybrid BNN Structure}
\begin{figure}
\centering
\includegraphics[width=0.47\textwidth]{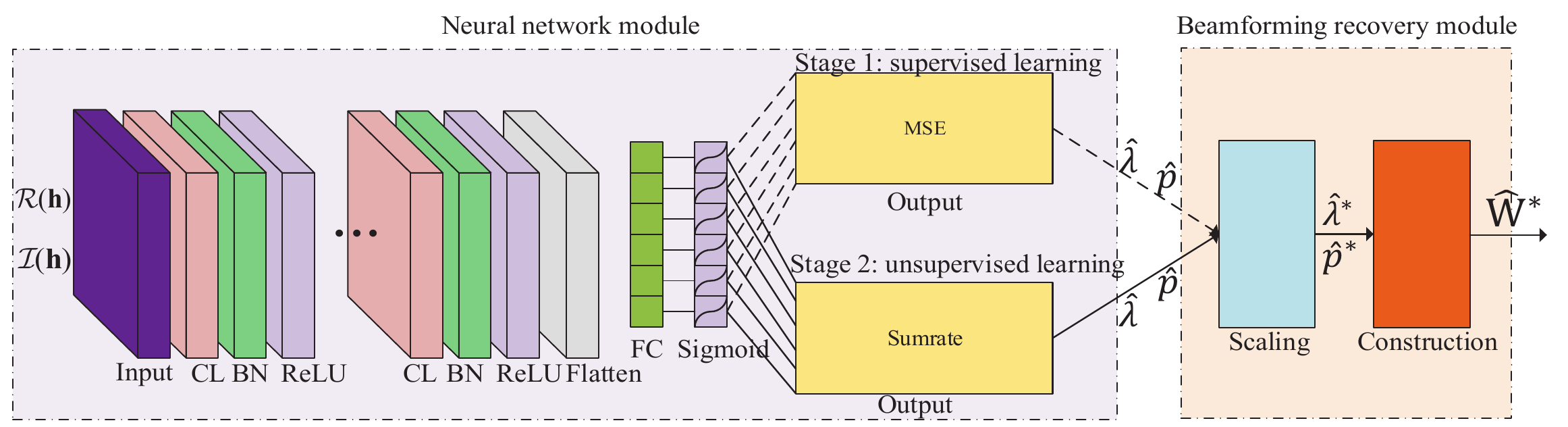}
\caption{BNN for the sum rate maximization problem.}
\label{BNN structure for sumrate}
\end{figure}
The BNN for problem \textbf{P3} is presented in Fig. \ref{BNN structure for sumrate}. The major difference from the BNNs in Figs. \ref{BNN structure for sinr} and \ref{BNN structure for power}  is that the BNN in Fig. \ref{BNN structure for sumrate} has two stages of training. The first stage is responsible for pre-training using the supervised learning method with the loss function based on the MSE metric \eqref{loss1}, while the second stage is responsible for enhanced training using the unsupervised learning method with the loss function whose metric is the objective function \eqref{loss2}. Such a hybrid learning method of the supervised and unsupervised learning can significantly improve the learning performance and also accelerate convergence \cite{lee2018deep}.  More specifically, the pre-training, as the approximation of WMMSE algorithm, starts with the  random initialization of neural network parameters and the loss function \eqref{loss1}. After the pre-training is finished, the neural network parameters are reserved   and the loss function is replaced by \eqref{loss2}, such that the second-stage training can achieve improved performance than the first-stage training.

Different from the  BNNs in Figs. \ref{BNN structure for sinr}  and  \ref{BNN structure for power},  the output layer in Fig. \ref{BNN structure for sumrate} generates $2K$ values including the power allocation vectors $\hat{\qp}$ and $\hat{\bm{\lambda}}$. Then the scaling layer scales the results of the output layer $\hat{\qq}$ and  $\hat{\bm{\lambda}}$ to meet the power constraint by the following method:
\begin{equation}
\hat{\qp}^{\ast}=\frac{P_{max}}{||\hat{\qp}||_1}\hat{\qp} \ \text{and} \ \hat{\bm{\lambda}}^{\ast}=\frac{P_{max}}{||\hat{\bm{\lambda}}||_1}\hat{\bm{\lambda}}.
\end{equation}
Finally, the construction layer constructs the downlink beamforming vectors according to \eqref{solution struc of sumrate}:
\begin{equation}
  \hat{\qw}_k^{\ast}=\sqrt{\hat{p}^{\ast}_k}\frac{(\qI_N+\sum^K_{k=1}{\frac{\hat{\lambda}^{\ast}_k}{\sigma^2}\qh_k\qh_k^H})^{-1}\qh_k}{||(\qI_N+\sum^K_{k=1}{\frac{\hat{\lambda}^{\ast}_k}{\sigma^2}\qh_k\qh_k^H})^{-1}\qh_k||_2}, \forall k.
\end{equation}
Thus, the time complexity of the beamforming recovery module for problem \textbf{P3} is $\mathcal{O}(KN^2+N^3)$.
\section{Simulation Results}\label{section simulation results}

To evaluate the performance of the proposed BNNs, we carry out numerical simulations to compare the BNNs with several benchmark solutions (when available), including the optimal beamforming, the ZF beamforming  \cite{Joham2005Linear}, the RZF beamforming  \cite{spencer2004zero}, and the WMMSE algorithm. We consider a downlink transmission scenario where the BS is equipped with $N=6$ antennas and  its coverage is a disc with a radius of 500 m. There are $K=4$ single-antenna users and these users are distributed uniformly within the coverage of the BS. Note that none of these users is closer to the BS than 100 m. The channel of user $k$ is modelled as $\qh_k=\sqrt{d_k}\tilde{\qh}_k\in\mathbb{C}^{N\times 1}$ where $\tilde{\qh}_k\sim\mathcal{CN}(\bm{0},\mathbf{I}_N)$  is the  small-scale fading \cite{zhu2012chunk} and $d_k=128.1 + 37.6\log_{10}(\omega)$[dB] denotes the pathloss between user $k$ and the BS  \cite{dahrouj2010coordinated} with $\omega$ representing the distance in km. Here, shadow fading is omitted for simplicity. The noise power spectral density is $-174$ dBm/Hz and the total system bandwidth is 20 MHz.  For simplicity, we assume all the sub-streams have the same importance and all the users  have  the same priority, i.e., $\rho_k=1, \forall k,$ and $\alpha_k=1, \forall k$. Besides, perfect CSI is assumed to be available at the BS.

In our simulation, we prepare 20000 training samples and 5000 testing samples, respectively. The validation split is set to 0.2 and the training data is randomly shuffled at each epoch.  All the  BNNs have the same structure as shown in Table \ref{BNN para}.  The fully-connected layer in the BNNs for problems \textbf{P1} and \textbf{P2} has $K$ neurons but that in the BNN for problem \textbf{P3} has $2K$ neurons. 
The Glorot normal initializer \cite{glorot2010understanding} is used for weight initialization and biases are initialized to 0.   Adam optimizer \cite{ba2015adam} is used with the MSE metric-based loss function. However, in the second stage of the BNN for problem \textbf{P3}, the metric of the loss function becomes the sum rate. The last activation layer is  the sigmoid function so that the target output in the training and testing samples should be normalized into (0,1] by dividing a factor. Also, the channel coefficients are normalized by the noise power before being fed into the BNNs to avoid entering the insensitive area of the sigmoid function. 
Note that unless explicitly mentioned otherwise, all the neural network modules adopt the default setting in Table I and  a separate neural network model is trained for  each different case.
\begin{table}[tbp]
\centering  %
\caption{Parameters of the neural network modules.}
\setlength{\tabcolsep}{1.5mm}{\begin{tabular}{ll}
\hline
Layer& Parameter\\
\hline
Layer 1 (input)& \begin{minipage}{4cm} \vspace{0.1cm}Input of size $2\times NK$, batch  of size 200, 100 epochs \vspace{0.1cm}\end{minipage} \\
Layer 2 (convolutional)&  \begin{minipage}{4cm} \vspace{0.1cm}8 kernels of $3\times 3$, zero padding  1, stride 1\vspace{0.1cm}\end{minipage} \\
Layer 3 (batch normalization)& Momentum=0.99, $\epsilon=0.001$\\
Layer 4 (activation)& ReLU\\
Layer 5 (convolutional) & \begin{minipage}{4cm} \vspace{0.1cm}8 kernels of $3\times 3$, zero padding  1, stride 1\vspace{0.1cm}\end{minipage} \\
Layer 7 (batch normalization) & Momentum=0.99, $\epsilon=0.001$\\
Layer 6  (activation)  & ReLU\\
Layer 8 (flatten) \\
Layer 9 (fully-connected)& $K$ or $2K$ neurons\\
Layer 10 (activation)  & Sigmoid\\
Layer 11 output layer &  \begin{minipage}{4cm} \vspace{0.1cm}Adam optimizer, learning rate of 0.001, MSE metric\vspace{0.1cm}\end{minipage} \\
\hline
\end{tabular}}
\label{BNN para}
\end{table}

\subsection{BNN  for the SINR Balancing Problem}

\begin{figure}
  \centering
  \subfigure[]{
    \label{sinr_verus_snr} 
    \includegraphics[width=0.47\textwidth]{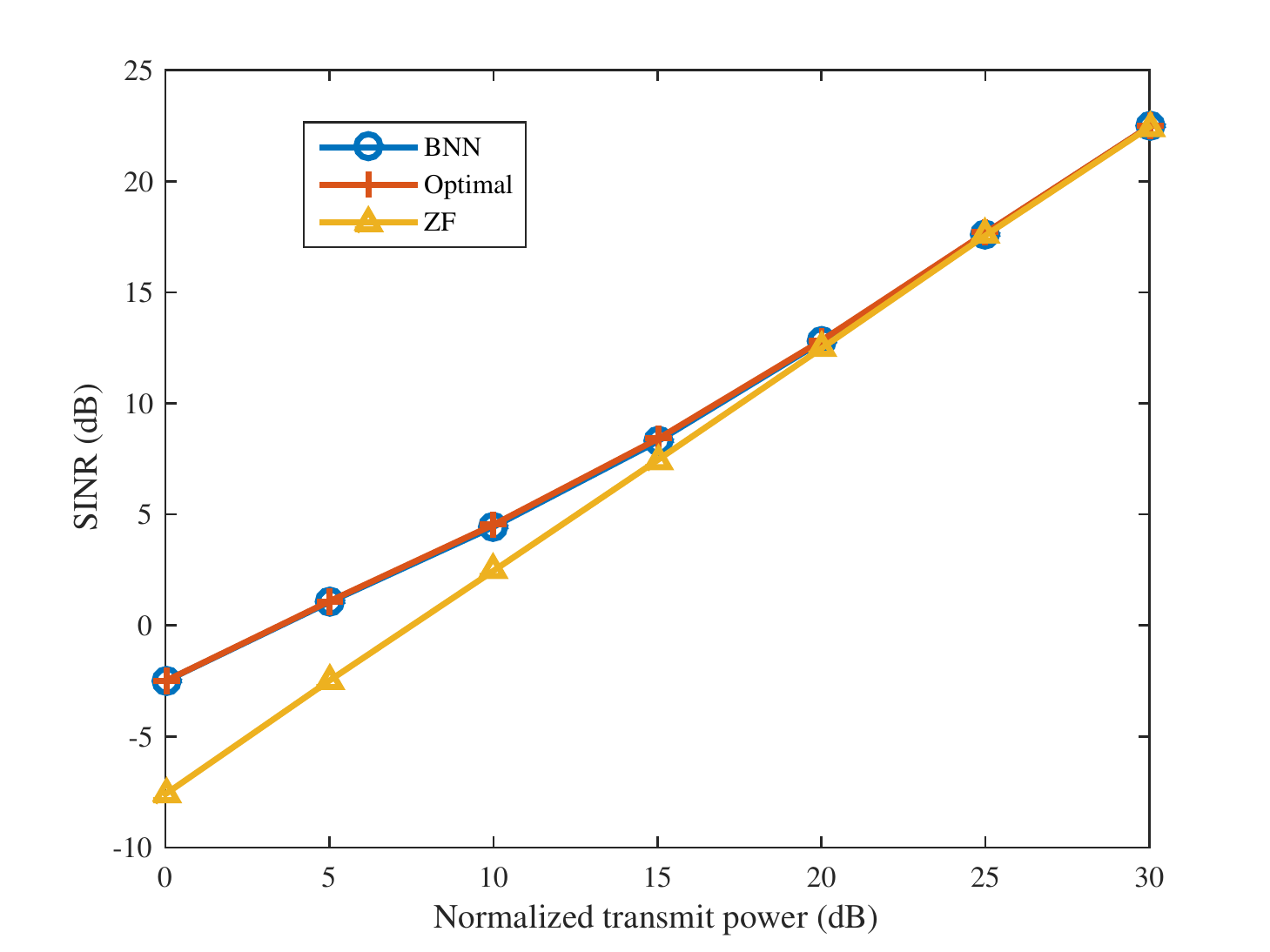}}
  \subfigure[]{
    \label{sinr_versus_power} 
    \includegraphics[width=0.47\textwidth]{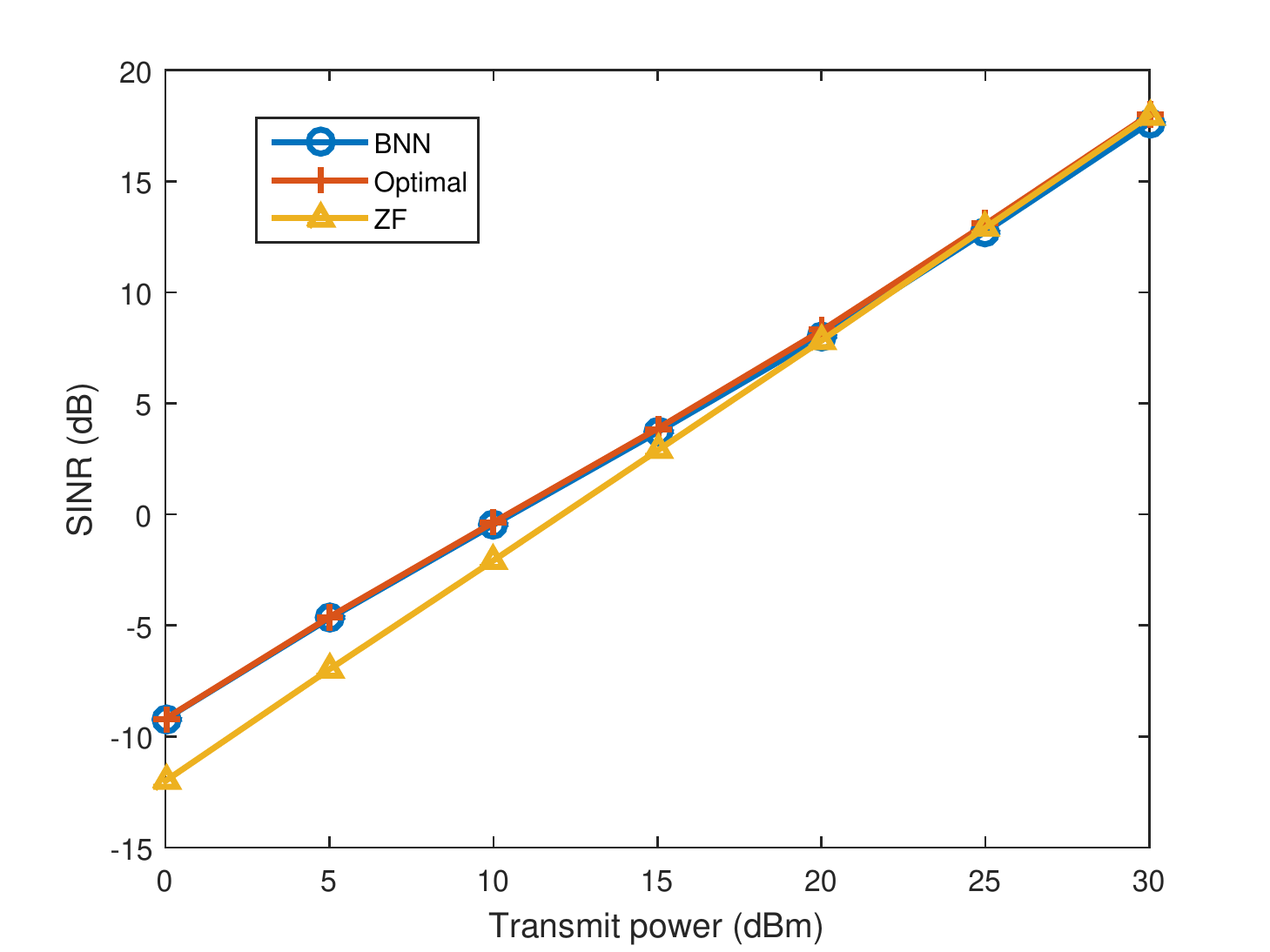}}
  \caption{The SINR performance averaged over 5000 samples  in two different cases: (a) without large-scale fading and (b) with large-scale fading under \{$K=4$, $N=6$\}.} 
  \label{sinr_over_power}
\end{figure}

We first consider the BNN for the SINR balancing problem \textbf{P1}, which updates network parameters in a supervised learning way. The iterative algorithm in \cite[Table 1]{schubert2004solution} is used to generate the training and testing samples. The ZF beamforming is achieved by allocating power to make all the users have the same SINR value under a total power constraint. Fig. \ref{sinr_over_power} shows the SINR performance averaged over 5000 samples in  two cases:  one only considering the small-scale fading but the other considering both the small-scale fading and large-scale fading. In both cases, the SINR performance of the proposed BNN solution is very close to that of the optimal solution \cite{schubert2004solution}. It is observed that there is an obvious gap between the optimal solution and the ZF beamforming in the low normalized transmit-power ($\frac{P_{max}}{\sigma^2}$) regime of Fig. \ref{sinr_verus_snr} as well  as the low transmit-power regime of Fig. \ref{sinr_versus_power}. However, the gap decreases  as the (normalized) transmit power increases.

\begin{figure}
\centering
\includegraphics[width=0.47\textwidth]{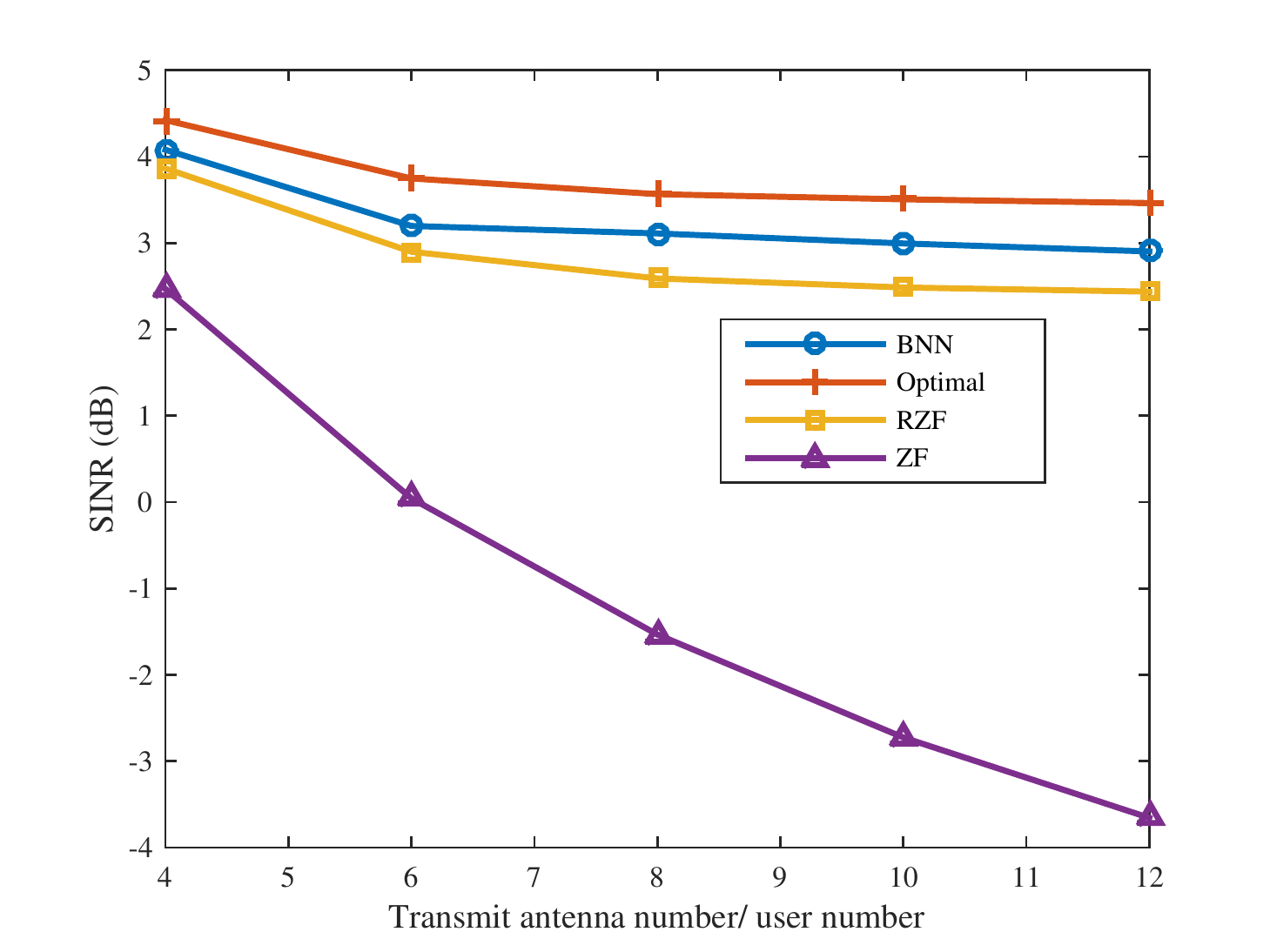}
\caption{Comparison of four different beamforming solutions, i.e., the optimal solution, the ZF beamforming, the RZF beamforming, and the BNN solution under \{$K=N$, $P_{max}=20$ dBm\}.}
\label{sinr_versus_antennaNum}
\end{figure}
\begin{figure}
\centering
\includegraphics[width=0.47\textwidth]{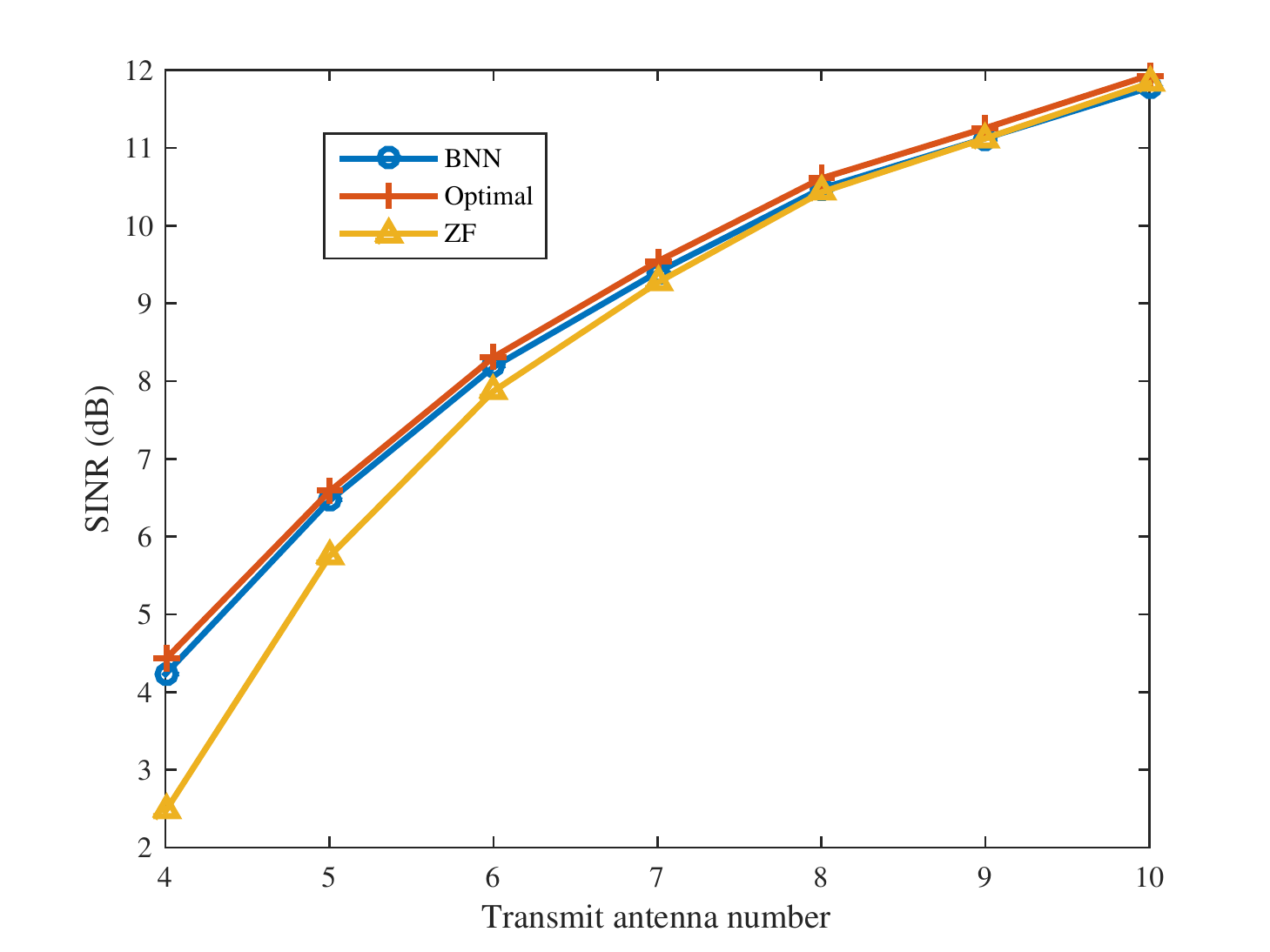}
\caption{The SINR performance versus different transmit antenna numbers  using the same trained BNN under \{$K=4$, $N=10$, $P_{max}=20$ dBm\}.}
\label{sinr_generality}
\end{figure}

To further compare the SINR performance of the optimal solution, the ZF beamforming, the RZF beamforming whose regularization parameter is set as $\frac{P_{max}}{K}$, and the BNN solution,  we evaluate the output SINR in Fig. \ref{sinr_versus_antennaNum} assuming that the number of users  is the same as the number of BS antennas, i.e., $K=N$, and they increase together. It is shown that the BNN solution has some performance loss compared to the optimal solution due to the estimation error,  but the BNN solution always achieves a better performance than the ZF beamforming and  RZF beamforming. This fact indicates the application prospect of the BNN: the computational complexity and time of the BNN solution is similar to those of the ZF beamforming and RZF beamforming,  but is much lower than that of the optimal solution  because the optimal solution relies on an iterative process.
Besides, we also find that the SINR performance of the four solutions decrease as the transmit antenna number (user number) increases and among the four solutions the ZF beamforming suffers most from the performance loss.

\begin{table}[tbp]
\centering  %
\caption{I/Q transformation versus P/M transformation.}
\setlength{\tabcolsep}{1.5mm}{\begin{tabular}{|c|c|ccccc|}
\hline
\multicolumn{2}{|c|}{K/N}& 4& 6 & 8&10&12\\
\hline
\multirow{2}*{I/Q transformation} &MSE& 0.084& 0.038& 0.022&0.014 & 0.010\\
\cline{2-7}
~&MAE& 0.223&0.147 & 0.111& 0.088&0.075\\
\hline
\multirow{2}*{P/M transformation} &MSE& 0.086&0.039&0.022&0.014&0.010\\
\cline{2-7}
&MAE& 0.225&0.149&0.111&0.087& 0.073\\
\hline
\end{tabular}}
\label{iq versus pm}
\end{table}

Table \ref{iq versus pm} presents the comparison of two input formats, i.e., \textbf{I/Q transformation} and \textbf{P/M transformation},  in terms of the MSE performance and MAE performance of the predicted normalized power under the case with $K=N$ and $P_{max}=20$ dBm. As shown in Table \ref{iq versus pm}, \textbf{I/Q transformation} and \textbf{P/M transformation} have close performance.

In Fig. \ref{sinr_generality}, we demonstrate the generality of the proposed BNN by fixing the user number as $K=4$ and the transmit power as $P_{max}=20$
dBm and show the SINR performance versus different transmit antenna settings. We train only a single BNN with \{$K=4$, $N=10$\}, but  allow the number of transmit antennas to vary from  4 to 10 when using the trained BNN. Then the redundant entries at the inputs and outputs are filled with 0's. It can be seen that  these predicted results are very close to that of the optimal solution. This fact suggests the generality of the BNN, i.e., we can train a large BNN with more antennas which will also work for the cases with less antennas without re-training. This will be useful when some transmit antennas of the BS are malfunctioning or turned off.

\subsection{BNN for the Power Minimization Problem}
\begin{figure}
  \centering
  \subfigure[]{
    \label{snr_feasibility_versus_sinr} 
    \includegraphics[width=0.47\textwidth]{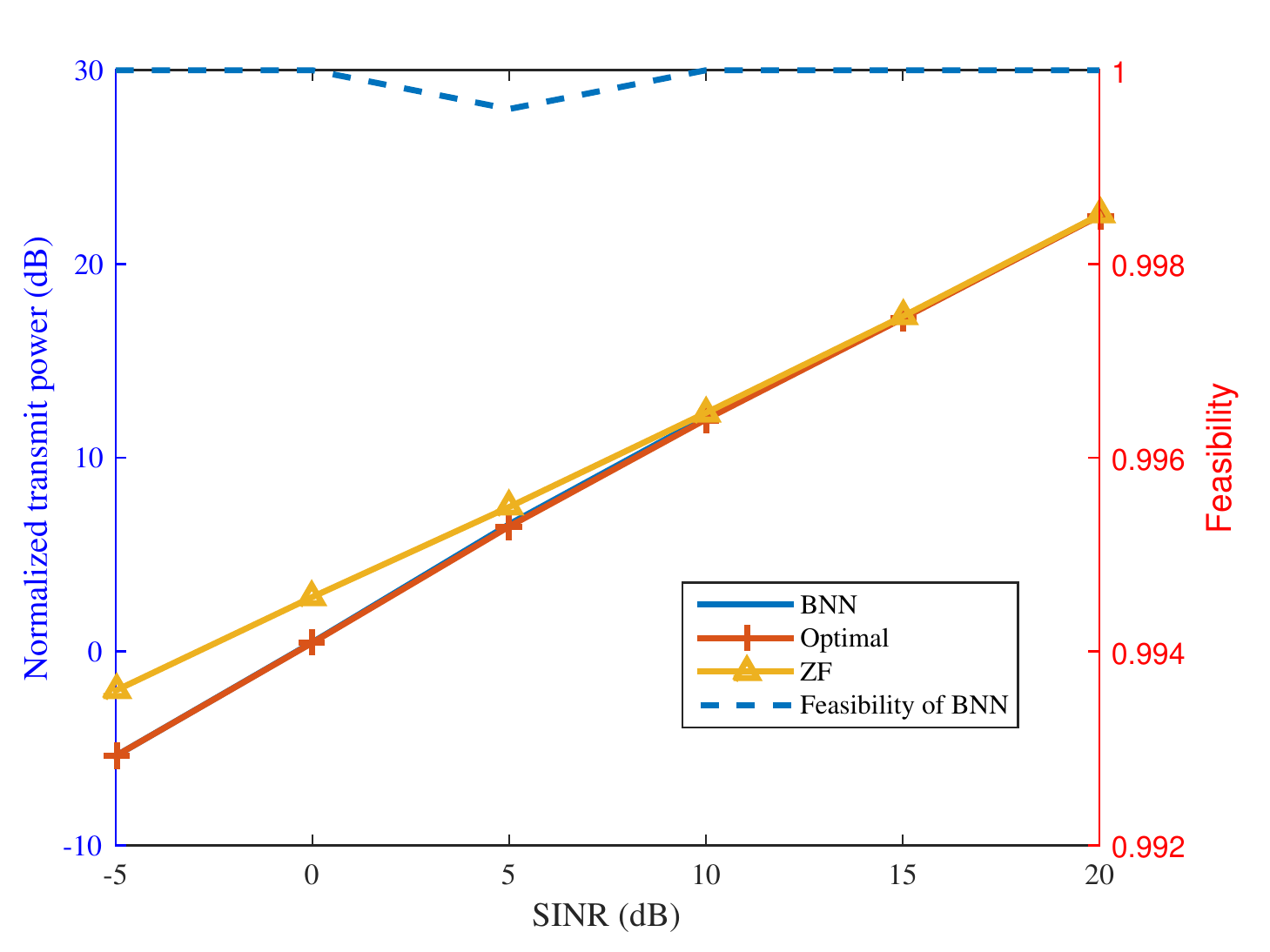}}
  \subfigure[]{
    \label{power_feasibility_versus_sinr} 
    \includegraphics[width=0.47\textwidth]{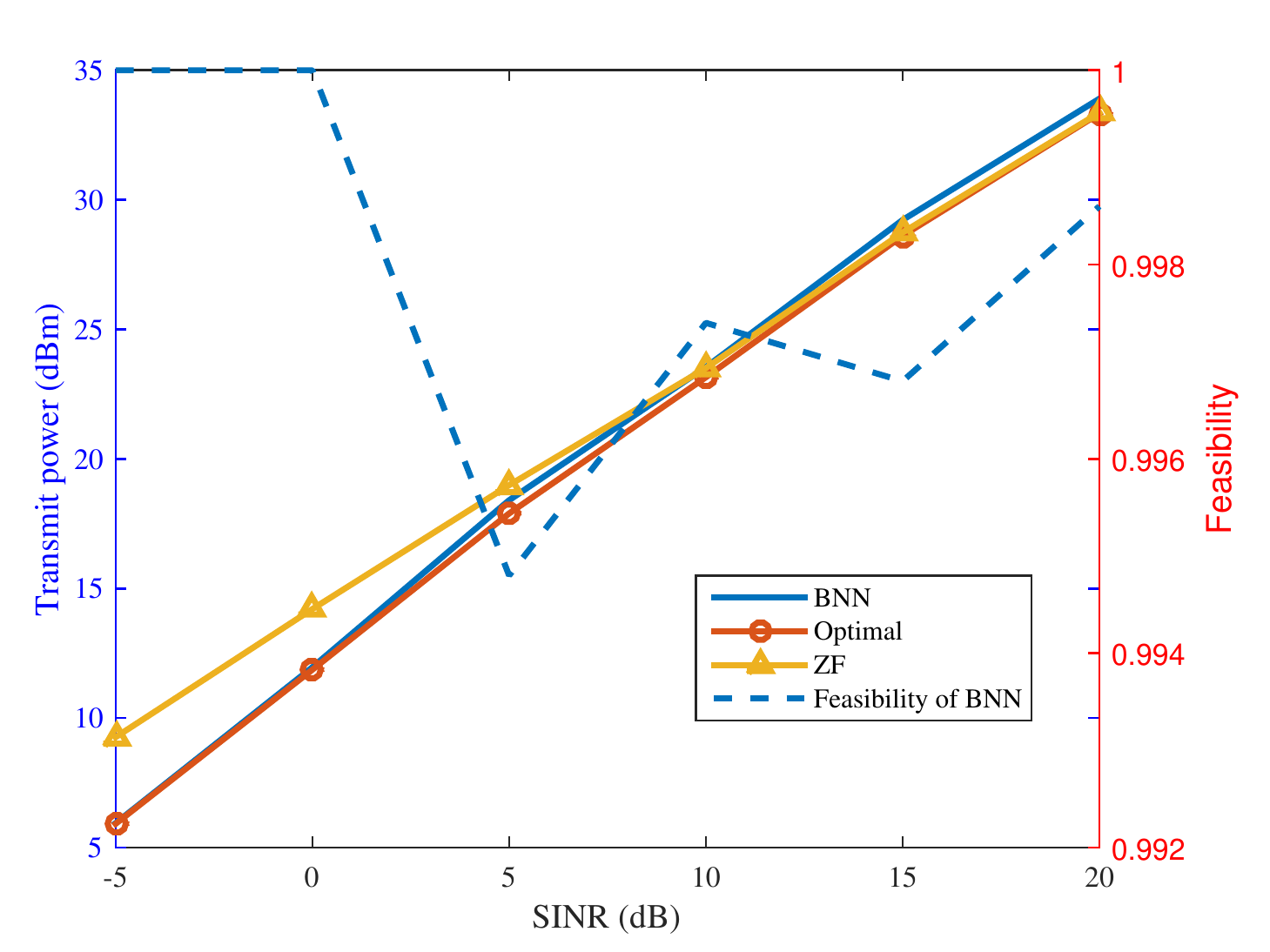}}
  \caption{The power performance averaged over the feasible sample set of the BNN solution  in two different cases: (a) without large-scale fading and (b) with large-scale fading under \{$K=4$, $N=6$\}.}
  \label{power_versus_sinr}

\end{figure}
In this subsection, we consider the BNN for the power minimization problem \textbf{P2}, which also updates network parameters  in a supervised learning way. The iterative algorithm in \cite{shi2016sinr} is used to generate the training and testing samples. The ZF beamforming for comparison is achieved by  minimizing the power for each user with a QoS constraint since there is no inter-user interference.  We first investigate the effect of  the SINR constraints of users on the power consumption. For convenience of comparison,  we assume the SINR constraints of all users are the same, i.e. $\Gamma_k=\Gamma, \forall k$.  In Fig. \ref{power_versus_sinr}, we compare the power performance of the optimal beamforming, the ZF beamforming, and the beamforming obtained by the BNN. Note that both  Figs. \ref{snr_feasibility_versus_sinr} and \ref{power_feasibility_versus_sinr} have two Y-axes  where the left Y-axis is used to measure the (normalized) transmit power averaged over the feasible sample set of the BNN solution  and the right Y-axis is used to show the feasibility of the BNN. As mentioned in Section \ref{section bnn for power minimization problem}, the BNN may fail to find a feasible solution to problem \textbf{P2} if the prediction error is unacceptable.

Figs. \ref{snr_feasibility_versus_sinr} and \ref{power_feasibility_versus_sinr} present the (normalized) transmit power performance  in the cases without and with consideration of the large-scale fading, respectively. In  both cases, the (normalized) transmit power performance of the BNN solution is close to that of the optimal solution, and significantly outperforms the ZF beamforming  in the low SINR-constraint regime which is higher than that of the optimal solution. We also find that, according to Fig. 8(b), the BNN solution performs slightly worse than the ZF solution when the SINR constraint is large, this is because the ZF solution becomes closer to the optimal solution as the SINR constraints increase, but the performance of the BNN solution is  still close to that of the optimal solution. This fact suggests that when the SINR constraints are high,  the ZF solution is a good choice instead of the BNN solution.
Besides, we find that the feasibility of the BNN solution in both cases is more than 99.4\%.

\begin{figure}
  \centering
  \subfigure[]{
    \label{power_versus_userNum} 
    \includegraphics[width=0.47\textwidth]{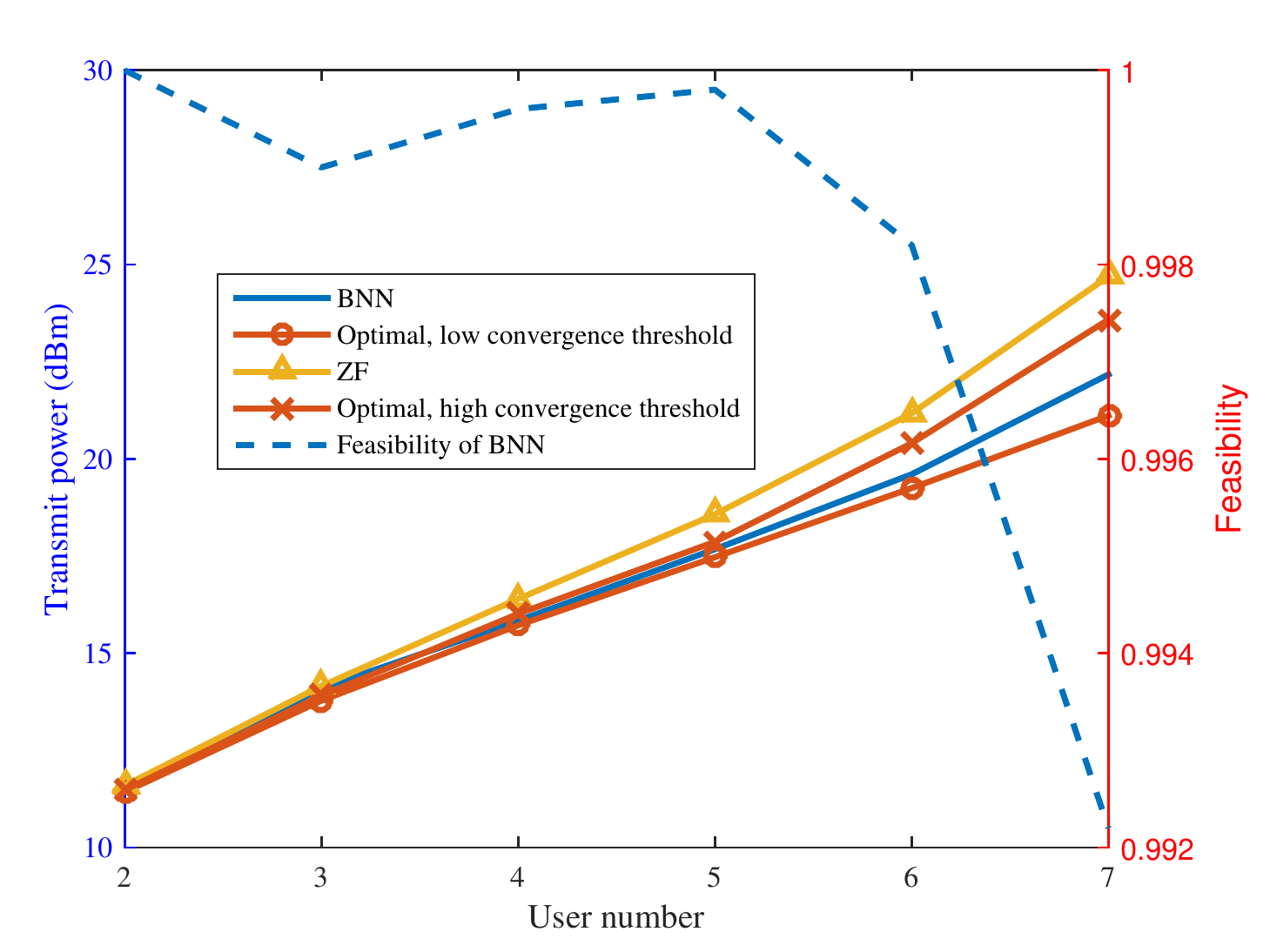}}
  \subfigure[]{
    \label{runningtime_versus_userNum} 
    \includegraphics[width=0.47\textwidth]{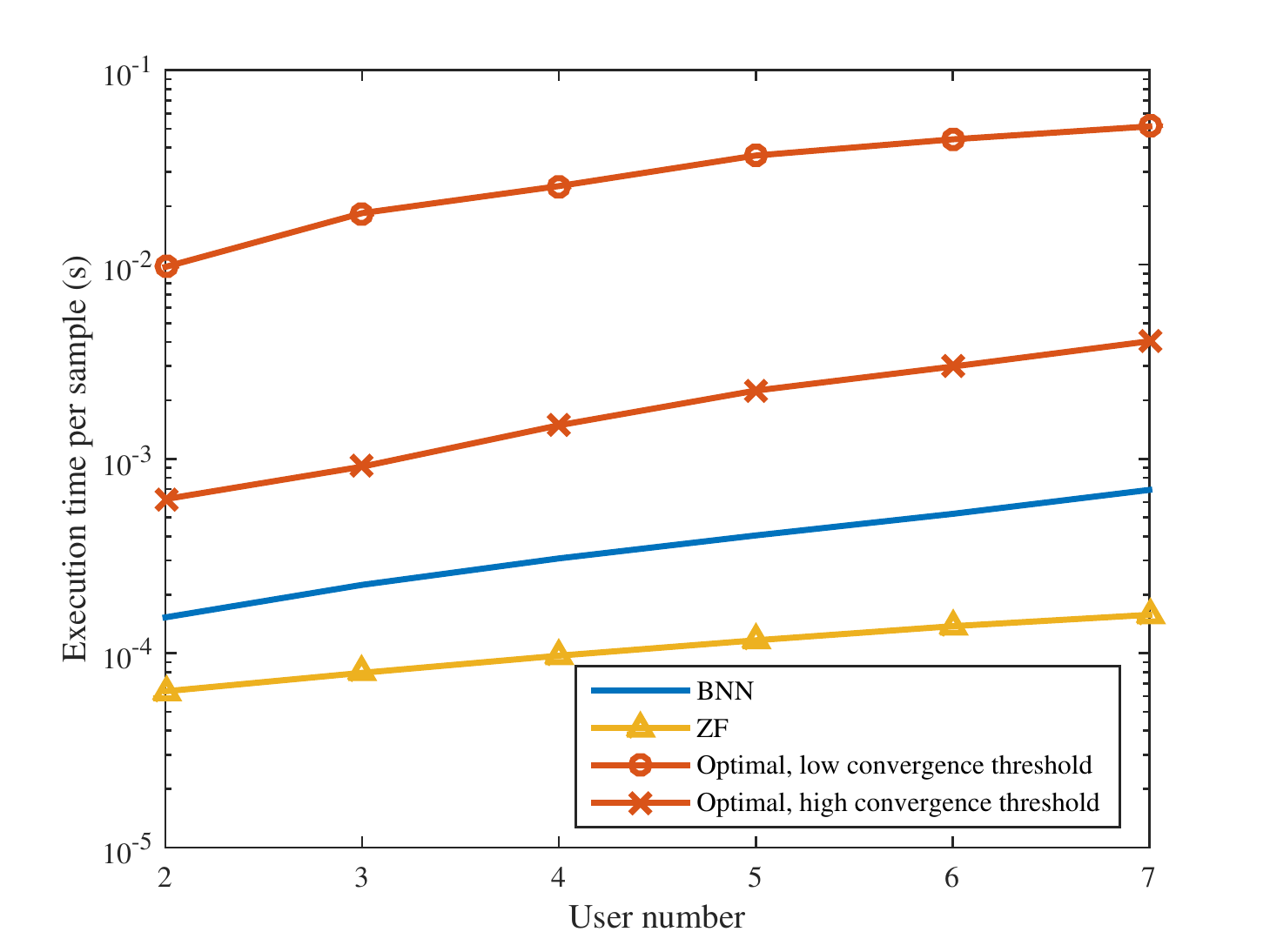}}
  \caption{Comparison of three different beamforming  solutions, i.e., the optimal solution, the BNN solution,  and ZF beamforming: (a) power performance  and (b) execution time per sample averaged over 5000 samples under \{$\Gamma=5$ dB, $N=8$\}.}
  \label{power_time_over_userNum}
\end{figure}
To further compare the BNN solution with the optimal solution and the  ZF beamforming, we plot their power performance  and execution time per sample in Figs. \ref{power_versus_userNum} and \ref{runningtime_versus_userNum}, respectively. Here, we consider two convergence strategies for the optimal iterative algorithm: the high convergence threshold ($\varepsilon_1=10^{-2}$) which can be reached with less iterations and the low convergence threshold ($\varepsilon_2=10^{-4}$) which requires more iterations for problem \textbf{P2}, i.e., $\frac{|\sum^K_{k=1}||\mathbf{w}^{(t-1)}_k||^2-\sum^K_{k=1}||\mathbf{w}^{(t)}_k||^2|}{\sum^K_{k=1}||\mathbf{w}^{(t-1)}_k||^2}\leq \varepsilon_\kappa,\kappa\in\{1,2\}$. In Fig. \ref{power_time_over_userNum}, the BS antenna number and SINR target of users are fixed as $N=8$ and $\Gamma=5$ dB. It is observed from Fig. \ref{power_versus_userNum} that as the user number $K$ increases,  the performance gap between the ZF beamforming and the optimal beamforming with the low convergence threshold becomes large because more users share the array gain.  The BNN solution, with the feasibility of up to 99\%, shows a better performance than the ZF beamforming and the optimal iterative algorithm with the high convergence threshold.  Fig. \ref{runningtime_versus_userNum} demonstrates that  compared to the optimal solution with the low convergence threshold, the BNN solution can reduce  the  execution time per sample by about two orders of magnitude,  which is slightly longer than that of the ZF beamforming. This is  because the BNN solution and the ZF beamforming are obtained without an iterative process, but the BNN needs to execute the neural network operations  as well as the conversion process. We can reduce the iteration times using the high convergence threshold, but this leads to the power performance degradation.   According to  the results in Figs. \ref{power_versus_userNum} and \ref{runningtime_versus_userNum}, we can conclude that the BNN solution provides a good balance between the performance and computational complexity.

\subsection{BNN for the Sum Rate Maximization Problem}
\begin{figure}
  \centering
  \subfigure[]{
    \label{sumrate_verus_snr} 
    \includegraphics[width=0.47\textwidth]{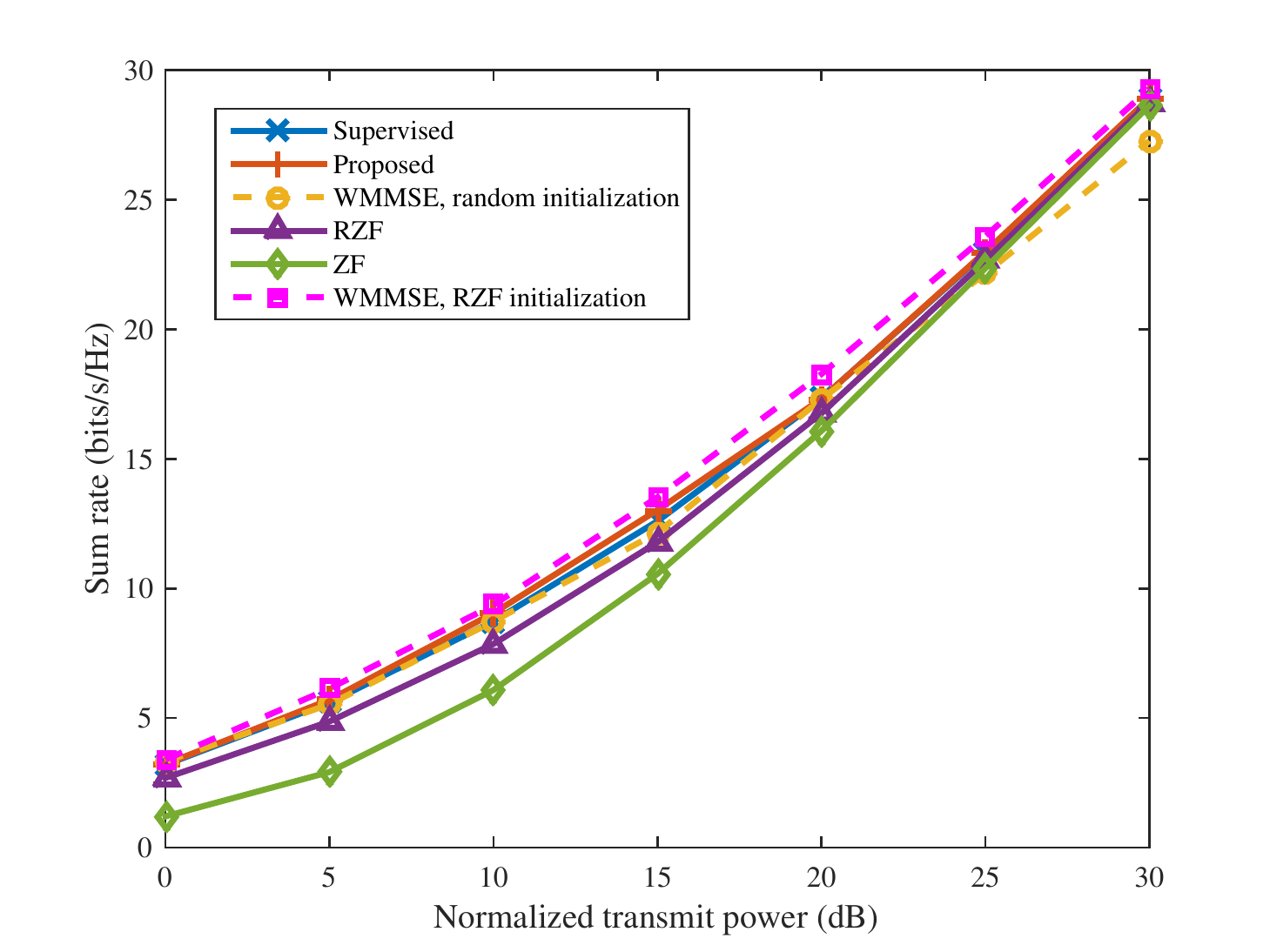}}
  \subfigure[]{
    \label{sumarate_versus_power} 
    \includegraphics[width=0.47\textwidth]{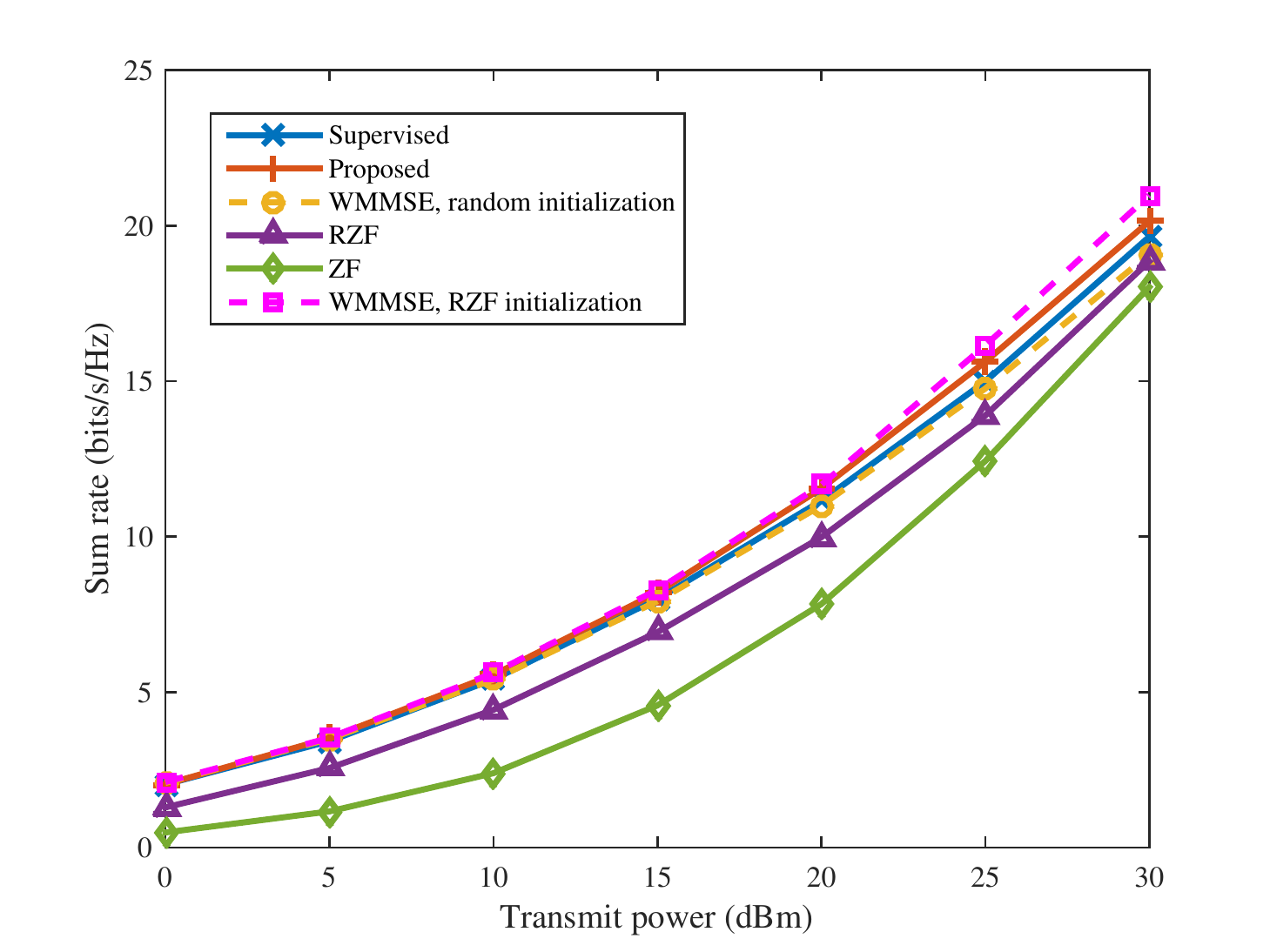}}
  \caption{The sum rate performance averaged over 5000 samples  in two different cases: (a) without large-scale fading and (b) with large-scale fading under \{$K=4$, $N=4$\}.}
  \label{sumrate_over_power}
\end{figure}
In this subsection, we evaluate the performance of the BNN for the sum rate maximization problem \textbf{P3} based on the proposed hybrid learning under the assumption that $K=4$ and  $N=4$. The ZF beamforming with $p_k=\frac{P_{max}}{K},\forall k$ and the RZF beamforming with $p_k=\lambda_k=\frac{P_{max}}{K},\forall k$ are introduced as two baseline solutions.  Since the performance of the WMMSE algorithm  heavily relies on initialization \cite{shi2011an,christensen2008weighted}, two different initialization methods, the RZF initialization and the random initialization, are considered and the WMMSE algorithm with the RZF initialization  is used  to generate samples for the supervised learning in the first stage. First, Fig. \ref{sumrate_over_power} shows the sum rate performance averaged over 5000 samples in two different cases: the former case  in  Fig. \ref{sumrate_verus_snr} only considers small-scale fading and  and the latter case in Fig. \ref{sumarate_versus_power} considers both small-scale fading and large-scale fading. It is shown that the sum rate performance of all solutions increases as the (normalized) transmit power increases and different initialization methods of the WMMSE algorithm have a large performance gap. We observe that in both cases the proposed BNN solution based on the hybrid learning always achieves a performance close to that of the WMMSE algorithm with the RZF initialization, while the performance of the supervised learning-based BNN solution is less satisfactory. This is because the second stage of the hybrid learning method aims to maximize the sum rate and its performance is bounded by the global optimal solution to problem \textbf{P3}. But the aim of the BNN solution based on the supervised learning is to achieve as close to the WMMSE solution as possible and its performance is restricted by the WMMSE solution, which  is verified in Figs. \ref{sumrate_verus_snr} and \ref{sumarate_versus_power}.
 \begin{figure}
  \centering
  \subfigure[]{
    \label{sumrate_versus_antennaNum}
    \includegraphics[width=0.47\textwidth]{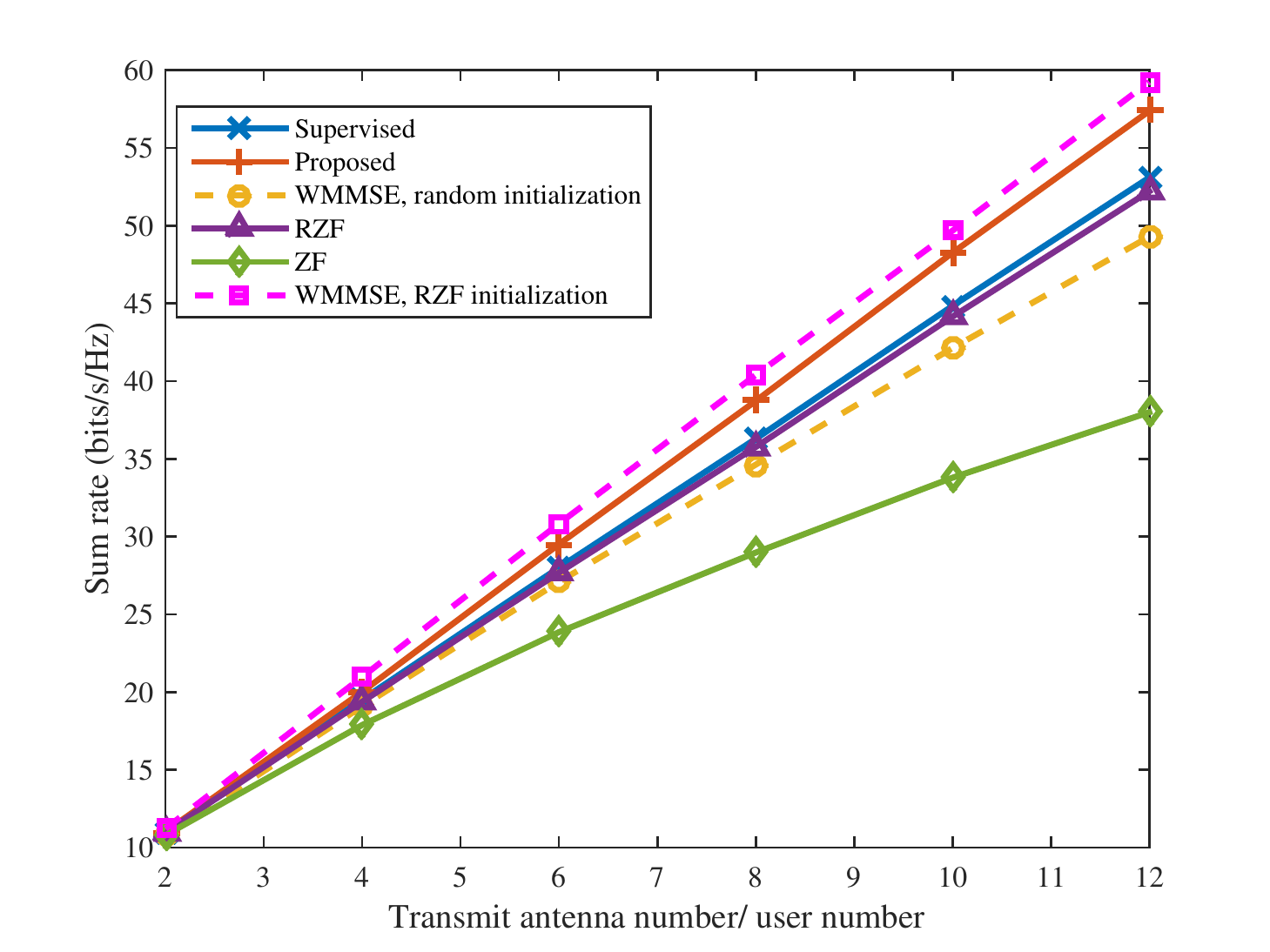}}
  \subfigure[]{
    \label{runningtime_versus_antennaNum}
    \includegraphics[width=0.47\textwidth]{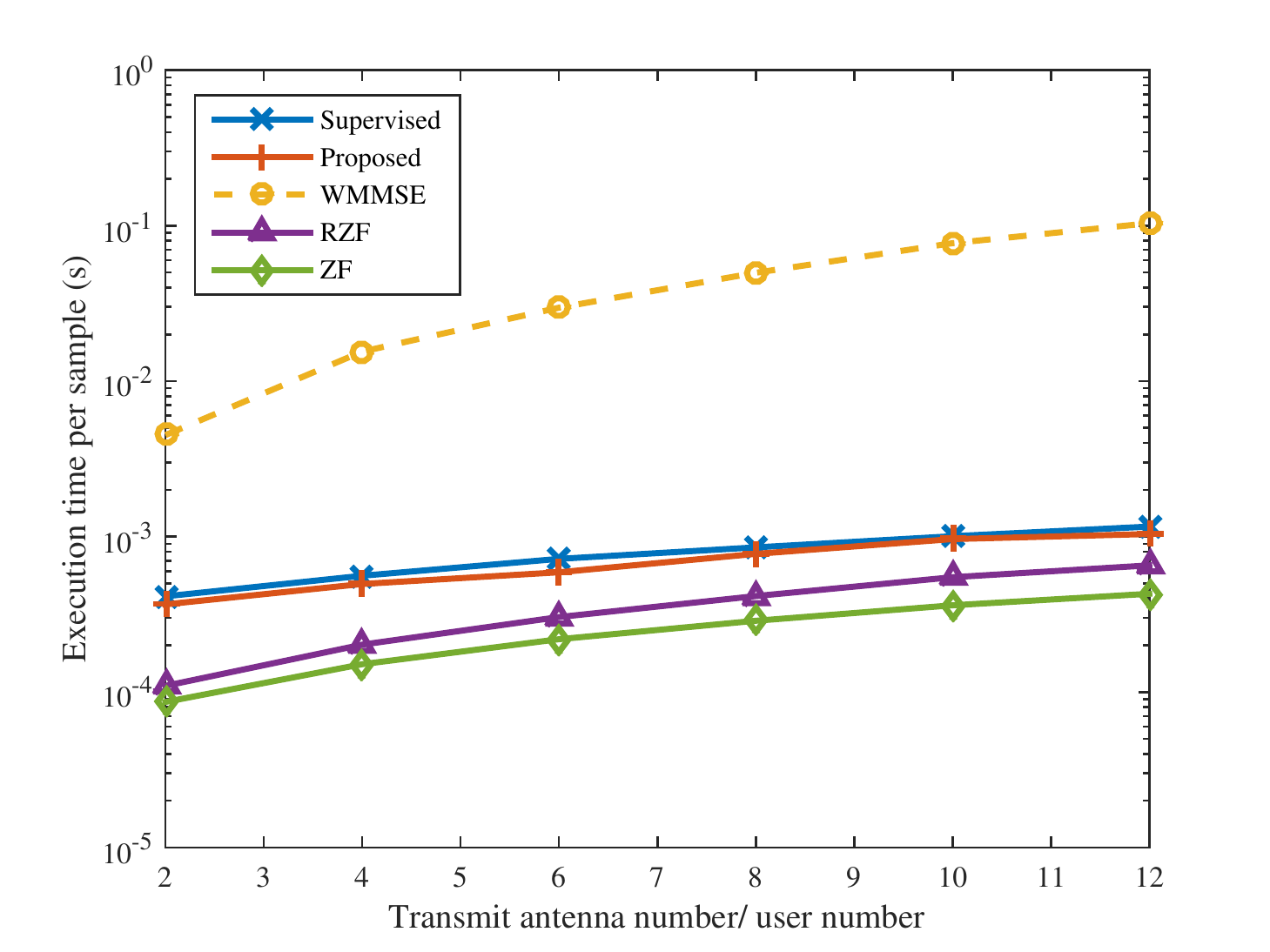}}
  \caption{Comparison of five different beamforming  solutions, i.e., the WMMSE solution, BNN solutions based on the supervised learning and the  proposed hybrid learning, respectively, the RZF beamforming, and the ZF beamforming: (a) sum rate performance  and (b) execution time per sample averaged over 5000 samples under \{$K=N$, $P_{max}=30$ dBm\}.}
  \label{sumrate_runningtime_versus_antennaNum}
\end{figure}

We further compare the sum rate performance and the computational complexity, in terms of the execution time per sample, of five beamforming solutions  in Figs. \ref{sumrate_versus_antennaNum} and \ref{runningtime_versus_antennaNum}, respectively. The iteration number of the WMMSE algorithm is limited to at most 10.  We fix the transmit power budget as $P_{max}=30$ dBm and assume the transmit antenna number is the same as the user number, i.e., $N=K$. As the number of transmit antennas  increases, the sum rate performance of all five solutions increases simultaneously. The performance of the proposed BNN solution based on the hybrid learning method is always  close to that of the WMMSE algorithm with the RZF initialization,  but is superior to those of the other four solutions and the performance gap becomes larger when the number of the transmit antenna increases. According to Fig. \ref{runningtime_versus_antennaNum},  the execution time per sample of the BNN solutions based on the supervised learning and hybrid learning methods is at the same level, which is slightly longer than that of the   ZF beamforming and the RZF beamforming, for the same reason of Fig. \ref{power_time_over_userNum}(b). As expected, the WMMSE algorithm consumes the most time because of its iterative process.  Similar to the other proposed BNNs, it proves that the proposed BNN solution to the sum rate problem \textbf{P3} provides a good balance between the performance and computational complexity.

\section{Conclusions}\label{section conclusion}
In this paper, we proposed a DL-based framework for fast optimization of the beamforming vectors in the MISO downlink  and then devised three BNNs under this framework for the SINR balancing problem under a total power constraint,  the power minimization problem under individual QoS constraints, and the sum rate maximization problem under a total power constraint, respectively. The proposed BNNs are based on the CNN structure and expert knowledge. The supervised learning method was adopted for  the SINR balancing problem and the power minimization problem because effective algorithms are available for generating training samples. However, there is no practically useful algorithm to find the optimal solution to the nonconvex sum rate maximization problem, therefore the corresponding BNN adoptes a hybrid learning method which first pre-trains the neural network based on the supervised learning method, and then updates the network parameters with the unsupervised learning method to further improve learning performance. Furthermore, in order to reduce the complexity of prediction, the proposed BNNs take advantage of expert knowledge to extract key features instead of predicting  beamforming matrix directly.  Simulation results demonstrated  that the proposed BNN solutions provided a good balance between the performance and  complexity, compared to the existing algorithms.

This work is an attempt  to apply the DL technique to beamforming optimization. Actually, a lot of extension works  are worth further study. For example, it is unclear so far which input format, I/Q transformation or P/M transformation, is better. In addition, the joint optimization of user selection and beamforming design for the power minimization problem is interesting and it deserves more investigation. {Besides,  user mobility, machine-type communications, imperfect CSI, feasibility detection, and multi-cell scenarios are also interesting extensions for future works.}

\section*{Acknowledgment}
We gratefully acknowledge the support of NVIDIA Corporation with the donation of the Titan Xp GPU used for this research.

\bibliographystyle{IEEEtran}

\begin{thebibliography}{10}
\providecommand{\url}[1]{#1}
\csname url@samestyle\endcsname
\providecommand{\newblock}{\relax}
\providecommand{\bibinfo}[2]{#2}
\providecommand{\BIBentrySTDinterwordspacing}{\spaceskip=0pt\relax}
\providecommand{\BIBentryALTinterwordstretchfactor}{4}
\providecommand{\BIBentryALTinterwordspacing}{\spaceskip=\fontdimen2\font plus
\BIBentryALTinterwordstretchfactor\fontdimen3\font minus
  \fontdimen4\font\relax}
\providecommand{\BIBforeignlanguage}[2]{{%
\expandafter\ifx\csname l@#1\endcsname\relax
\typeout{** WARNING: IEEEtran.bst: No hyphenation pattern has been}%
\typeout{** loaded for the language `#1'. Using the pattern for}%
\typeout{** the default language instead.}%
\else
\language=\csname l@#1\endcsname
\fi
#2}}
\providecommand{\BIBdecl}{\relax}
\BIBdecl

\bibitem{xia2018deep}
W.~Xia, G.~Zheng, Y.~Zhu, J.~Zhang, J.~Wang, and A.~Petropulu, ``Deep learning
  based beamforming neural networks in downlink {MISO} systems,'' in
  \emph{Proc. IEEE Int. Conf. Commun. (ICC) Workshop}, Shanghai, China, May
  2019, pp. 1--5.

\bibitem{bjornson2014optimal}
E.~Bj\"{o}rnson, M.~Bengtsson, and B.~Ottersten, ``Optimal multiuser transmit
  beamforming: {A} difficult problem with a simple solution structure,''
  \emph{IEEE Signal Process. Mag.}, vol.~31, no.~4, pp. 142--148, Jul. 2014.

\bibitem{boche2002general}
H.~Boche and M.~Schubert, ``A general duality theory for uplink and downlink
  beamforming,'' in \emph{Proc. IEEE Conf. Veh. Technol. Conf. (VTC)}, vol.~1,
  Vancouver, Canada, Sep. 2002, pp. 87--91.

\bibitem{gerlach1996base}
D.~Gerlach and A.~Paulraj, ``Base station transmitting antenna arrays for
  multipath environments,'' \emph{Signal Process.}, vol.~54, no.~1, pp. 59--73,
  Oct. 1996.

\bibitem{shi2016sinr}
Q.~Shi, M.~Razaviyayn, M.~Hong, and Z.~Luo, ``{SINR} constrained beamforming
  for a {MIMO} multi-user downlink system: {Algorithms} and convergence
  analysis,'' \emph{IEEE Trans. Signal Process.}, vol.~64, no.~11, pp.
  2920--2933, Jun. 2016.

\bibitem{rashid1998transmit}
F.~Rashid-Farrokhi, K.~R. Liu, and L.~Tassiulas, ``Transmit beamforming and
  power control for cellular wireless systems,'' \emph{IEEE J. Sel. Areas
  Commun.}, vol.~16, no.~8, pp. 1437--1450, Oct. 1998.

\bibitem{gershman2010convex}
A.~B. Gershman, N.~D. Sidiropoulos, S.~Shahbazpanahi, M.~Bengtsson, and
  B.~Ottersten, ``Convex optimization-based beamforming,'' \emph{IEEE Signal
  Process. Mag.}, vol.~27, no.~3, pp. 62--75, May 2010.

\bibitem{wiesel2006linear}
A.~Wiesel, Y.~C. Eldar, and S.~Shamai, ``Linear precoding via conic
  optimization for fixed {MIMO} receivers,'' \emph{IEEE Trans. Signal
  Process.}, vol.~54, no.~1, pp. 161--176, Jan. 2006.

\bibitem{shi2011an}
Q.~Shi, M.~Razaviyayn, Z.~Luo, and C.~He, ``An iteratively weighted {MMSE}
  approach to distributed sum-utility maximization for a {MIMO} interfering
  broadcast channel,'' \emph{IEEE Trans. Signal Process.}, vol.~59, no.~9, pp.
  4331--4340, Sep. 2011.

\bibitem{christensen2008weighted}
S.~S. Christensen, R.~Agarwal, E.~D. Carvalho, and J.~M. Cioffi, ``Weighted
  sum-rate maximization using weighted {MMSE} for {MIMO-BC} beamforming
  design,'' \emph{IEEE Trans. Wireless Commun.}, vol.~7, no.~12, pp.
  4792--4799, Dec. 2008.

\bibitem{Yoo2006on}
T.~Yoo and A.~Goldsmith, ``On the optimality of multiantenna broadcast
  scheduling using zero-forcing beamforming,'' \emph{IEEE J. Sel. Areas
  Commun.}, vol.~24, no.~3, pp. 528--541, Mar. 2006.

\bibitem{schubert2004solution}
M.~Schubert and H.~Boche, ``Solution of the multiuser downlink beamforming
  problem with individual {SINR} constraints,'' \emph{IEEE Trans. Veh.
  Technol.}, vol.~53, no.~1, pp. 18--28, Jan. 2004.

\bibitem{luo2010semidefinite}
Z.-Q. Luo, W.-K. Ma, A.~M.-C. So, Y.~Ye, and S.~Zhang, ``Semidefinite
  relaxation of quadratic optimization problems,'' \emph{IEEE Signal Process.
  Mag.}, vol.~27, no.~3, pp. 20--34, May 2010.

\bibitem{bengtesson2001optimal}
M.~Bengtsson and B.~Ottersten, ``Optimal and suboptimal transmit beamforming,''
  \emph{in Handbook of Antennas in Wireless Communications}, CRC Press, Jan.
  2001.

\bibitem{cvx}
M.~Grant and S.~Boyd, ``{CVX}: Matlab software for disciplined convex
  programming, version 2.1,'' \url{http://cvxr.com/cvx}, Mar. 2014.

\bibitem{rashid1998joint}
F.~Rashid-Farrokhi, L.~Tassiulas, and K.~R. Liu, ``Joint optimal power control
  and beamforming in wireless networks using antenna arrays,'' \emph{IEEE
  Trans. Commun.}, vol.~46, no.~10, pp. 1313--1324, Oct. 1998.

\bibitem{nachmani2016learning}
E.~Nachmani, Y.~Be'ery, and D.~Burshtein, ``Learning to decode linear codes
  using deep learning,'' in \emph{Proc. IEEE Annual Allerton Conf. Commun.
  Control Comput.}, Monticello, USA, Sep. 2016, pp. 341--346.

\bibitem{liang2018an}
F.~Liang, C.~Shen, and F.~Wu, ``An iterative {BP-CNN} architecture for channel
  decoding,'' \emph{IEEE J. Sel. Topics Signal Process.}, vol.~12, no.~1, pp.
  144--159, Feb. 2018.

\bibitem{fan2019cnn}
C.~Fan, X.~Yuan, and Y.-J.~A. Zhang, ``{CNN}-based signal detection for banded
  linear systems,'' \emph{IEEE Trans. Wireless Commun.}, vol.~18, no.~9, pp.
  4394 -- 4407, Sep. 2019.

\bibitem{samuel2019learning}
N.~{Samuel}, T.~{Diskin}, and A.~{Wiesel}, ``Learning to detect,'' \emph{IEEE
  Trans. Signal Process.}, vol.~67, no.~10, pp. 2554--2564, May 2019.

\bibitem{farsad2017detection}
N.~Farsad and A.~Goldsmith, ``Detection algorithms for communication systems
  using deep learning,'' \emph{arXiv preprint arXiv:1705.08044}, 2017.

\bibitem{wen2018deep}
C.-K. Wen, W.-T. Shih, and S.~Jin, ``Deep learning for massive {MIMO CSI}
  feedback,'' \emph{IEEE Wireless Commun. Lett.}, vol.~7, no.~5, pp. 748--751,
  Oct. 2018.

\bibitem{wang2019deep}
T.~Wang, C.-K. Wen, S.~Jin, and G.~Y. Li, ``Deep learning-based {CSI} feedback
  approach for time-varying massive {MIMO} channels,'' \emph{IEEE Wireless
  Commun. Lett.}, vol.~8, no.~2, pp. 416--419, Apr. 2019.

\bibitem{ye2018power}
H.~Ye, G.~Y. Li, and B.~Juang, ``Power of deep learning for channel estimation
  and signal detection in {OFDM} systems,'' \emph{IEEE Wireless Commun. Lett.},
  vol.~7, no.~1, pp. 114--117, Feb. 2018.

\bibitem{eisen2019learning}
M.~{Eisen}, C.~{Zhang}, L.~F.~O. {Chamon}, D.~D. {Lee}, and A.~{Ribeiro},
  ``Learning optimal resource allocations in wireless systems,'' \emph{IEEE
  Trans. Signal Process.}, vol.~67, no.~10, pp. 2775--2790, May 2019.

\bibitem{ahmed2019deep}
K.~I. {Ahmed}, H.~{Tabassum}, and E.~{Hossain}, ``Deep learning for radio
  resource allocation in multi-cell networks,'' \emph{IEEE Network}, pp. 1--8,
  2019.

\bibitem{sun2017learning}
H.~Sun, X.~Chen, Q.~Shi, M.~Hong, X.~Fu, and N.~D. Sidiropoulos, ``Learning to
  optimize: Training deep neural networks for wireless resource management,''
  in \emph{Proc. IEEE Int. Workshop Signal Process. Advances Wireless Commun.
  (SPAWC)}, Sapporo, Japan, Jul. 2017, pp. 1--6.

\bibitem{liang2018towards}
F.~{Liang}, C.~{Shen}, W.~{Yu}, and F.~{Wu}, ``Towards optimal power control
  via ensembling deep neural networks,'' \emph{IEEE Trans. Commun.}, pp. 1--1,
  2019.

\bibitem{lee2018deep}
W.~Lee, M.~Kim, and D.-H. Cho, ``Deep power control: Transmit power control
  scheme based on convolutional neural network,'' \emph{IEEE Commun. Lett.},
  vol.~22, no.~6, pp. 1276--1279, Jun. 2018.

\bibitem{van2019power}
T.~Van~Chien, T.~N. Canh, E.~Bj{\"o}rnson, and E.~G. Larsson, ``Power control
  in cellular massive {MIMO} with varying user activity: A deep learning
  solution,'' \emph{arXiv preprint arXiv:1901.03620}, 2019.

\bibitem{sanguinetti2018deep}
L.~{Sanguinetti}, A.~{Zappone}, and M.~{Debbah}, ``Deep learning power
  allocation in massive {MIMO},'' in \emph{Proc. Asilomar Conf. Signals,
  Systems, Computers}, Pacific Grove, CA, USA, Oct. 2018, pp. 1257--1261.

\bibitem{chen2017echo}
M.~{Chen}, W.~{Saad}, C.~{Yin}, and M.~{Debbah}, ``Echo state networks for
  proactive caching in cloud-based radio access networks with mobile users,''
  \emph{IEEE Trans. Wireless Commun.}, vol.~16, no.~6, pp. 3520--3535, Jun.
  2017.

\bibitem{Dorner2018deep}
S.~D\"{o}rner, S.~Cammerer, J.~Hoydis, and S.~t.~Brink, ``Deep learning based
  communication over the air,'' \emph{IEEE J. Sel. Topics Signal Process.},
  vol.~12, no.~1, pp. 132--143, Feb. 2018.

\bibitem{oshea2016learning}
T.~J. O'Shea, K.~Karra, and T.~C. Clancy, ``Learning to communicate: Channel
  auto-encoders, domain specific regularizers, and attention,'' in \emph{Proc.
  IEEE Int. Symp. Signal Process. Inf. Technol. (ISSPIT)}, Limassol, Cyprus,
  Dec. 2016, pp. 223--228.

\bibitem{zhao2018deep}
Z.~Zhao, ``Deep-waveform: {A} learned {OFDM} receiver based on deep complex
  convolutional networks,'' \emph{arXiv preprint arXiv:1810.07181}, 2018.

\bibitem{zhang2019deep}
C.~{Zhang}, P.~{Patras}, and H.~{Haddadi}, ``Deep learning in mobile and
  wireless networking: {A} survey,'' \emph{IEEE Commun. Surveys Tutorials},
  vol.~21, no.~3, pp. 2224 -- 2287, third quarter 2019.

\bibitem{zappone2019wireless}
A.~{Zappone}, M.~{Di Renzo}, and M.~{Debbah}, ``Wireless networks design in the
  era of deep learning: Model-based, {AI}-based, or both?'' \emph{IEEE Trans.
  Commun.}, vol.~67, no.~10, pp. 7331 -- 7376, Oct. 2019.

\bibitem{wang2017deep}
T.~Wang, C.-K. Wen, H.~Wang, F.~Gao, T.~Jiang, and S.~Jin, ``Deep learning for
  wireless physical layer: {Opportunities} and challenges,'' \emph{China
  Commun.}, vol.~14, no.~11, pp. 92--111, Nov. 2017.

\bibitem{Kerret2018robust}
P.~de~Kerret and D.~Gesbert, ``Robust decentralized joint precoding using team
  deep neural network,'' in \emph{Proc. Int. Symp. Wireless Commun. Systems
  (ISWCS)}, Lisbon, Portugal, Aug. 2018.

\bibitem{alkhateeb2018deep}
A.~Alkhateeb, S.~Alex, P.~Varkey, Y.~Li, Q.~Qu, and D.~Tujkovic, ``Deep
  learning coordinated beamforming for highly-mobile millimeter wave systems,''
  \emph{IEEE Access}, vol.~6, pp. 37\,328--37\,348, 2018.

\bibitem{shi2018learning}
Y.~Shi, A.~Konar, N.~D. Sidiropoulos, X.~Mao, and Y.~Liu, ``Learning to
  beamform for minimum outage,'' \emph{IEEE Trans. Signal Process.}, vol.~66,
  no.~19, pp. 5180--5193, Oct. 2018.

\bibitem{huang2019unsupervised}
H.~Huang, W.~Xia, J.~Xiong, J.~Yang, G.~Zheng, and X.~Zhu, ``Unsupervised
  learning based fast beamforming design for downlink {MIMO},'' \emph{IEEE
  Access}, vol.~7, pp. 7599--7605, 2019.

\bibitem{hornik1989multilayer}
K.~Hornik, M.~Stinchcombe, and H.~White, ``Multilayer feedforward networks are
  universal approximators,'' \emph{Neural networks}, vol.~2, no.~5, pp.
  359--366, 1989.

\bibitem{zhou2019universality}
D.-X. Zhou, ``Universality of deep convolutional neural networks,''
  \emph{Applied Computational Harmonic Analysis}, Jun. 2019.

\bibitem{Leem2018deep}
M.~{Lee}, Y.~{Xiong}, G.~{Yu}, and G.~Y. {Li}, ``Deep neural networks for
  linear sum assignment problems,'' \emph{IEEE Wireless Commun. Lett.}, vol.~7,
  no.~6, pp. 962--965, Dec. 2018.

\bibitem{cui2019spatial}
W.~Cui, K.~Shen, and W.~Yu, ``Spatial deep learning for wireless scheduling,''
  \emph{IEEE J. Sel. Areas Commun.}, vol.~37, no.~6, pp. 1248--1261, Jun. 2019.

\bibitem{yu2007transmitter}
W.~Yu and T.~Lan, ``Transmitter optimization for the multi-antenna downlink
  with per-antenna power constraints,'' \emph{IEEE Trans. Signal Process.},
  vol.~55, no.~6, pp. 2646--2660, Jun. 2007.

\bibitem{kulin2018end}
M.~Kulin, T.~Kazaz, I.~Moerman, and E.~D. Poorter, ``End-to-end learning from
  spectrum data: {A} deep learning approach for wireless signal identification
  in spectrum monitoring applications,'' \emph{IEEE Access}, vol.~6, pp.
  18\,484--18\,501, 2018.

\bibitem{ioffe2015batch}
S.~Ioffe and C.~Szegedy, ``Batch normalization: {Accelerating} deep network
  training by reducing internal covariate shift,'' in \emph{Proc. Int. Conf.
  Machine Learning (ICML)}, Lille, France, Jul. 2015, pp. 448--456.

\bibitem{botchkarev2019performance}
A.~Botchkarev, ``A new typology design of performance metrics to measure errors
  in machine learning regression algorithms,'' \emph{Interdisciplinary J. Inf.,
  Knowledge \& Management}, vol.~14, pp. 45--76, 2019.

\bibitem{matthiesen2018deep}
B.~Matthiesen, A.~Zappone, E.~A. Jorswieck, and M.~Debbah, ``A globally optimal
  energy-efficient power control framework and its efficient implementation in
  wireless interference networks,'' \emph{arXiv preprint arXiv:1812.06920},
  2018.

\bibitem{he2015convolutional}
K.~{He} and J.~{Sun}, ``Convolutional neural networks at constrained time
  cost,'' in \emph{Proc. IEEE Conf. Computer Vision Pattern Recognition
  (CVPR)}, Boston, MA, USA, Jun. 2015, pp. 5353--5360.

\bibitem{yang1998optimal}
W.~Yang and G.~Xu, ``Optimal downlink power assignment for smart antenna
  systems,'' in \emph{Proc. IEEE Int. Conf. Acoustics, Speech Process.
  (ICASSP)}, vol.~6, Seattle, USA, May 1998.

\bibitem{seneta2006non}
E.~Seneta, \emph{Non-negative matrices and Markov chains}.\hskip 1em plus 0.5em
  minus 0.4em\relax Springer Science \& Business Media, 2006.

\bibitem{ortega1990numerical}
J.~M. Ortega, \emph{Numerical analysis: a second course}.\hskip 1em plus 0.5em
  minus 0.4em\relax SIAM, 1990.

\bibitem{Visotsky199optimum}
E.~{Visotsky} and U.~{Madhow}, ``Optimum beamforming using transmit antenna
  arrays,'' in \emph{Proc. IEEE Veh. Technol. Conf.}, vol.~1, Houston, TX, USA,
  May 1999, pp. 851--856.

\bibitem{bjornson2013optimal}
E.~Bj{\"o}rnson and E.~Jorswieck, ``Optimal resource allocation in coordinated
  multi-cell systems,'' \emph{Foundations and Trends{\textregistered} in
  Communications and Information Theory}, vol.~9, no. 2--3, pp. 113--381, 2013.

\bibitem{Joham2005Linear}
M.~{Joham}, W.~{Utschick}, and J.~A. {Nossek}, ``Linear transmit processing in
  {MIMO} communications systems,'' \emph{IEEE Trans. Sig. Process.}, vol.~53,
  no.~8, pp. 2700--2712, Aug. 2005.

\bibitem{spencer2004zero}
Q.~H. {Spencer}, A.~L. {Swindlehurst}, and M.~{Haardt}, ``Zero-forcing methods
  for downlink spatial multiplexing in multiuser {MIMO} channels,'' \emph{IEEE
  Trans. Signal Process.}, vol.~52, no.~2, pp. 461--471, Feb. 2004.

\bibitem{zhu2012chunk}
H.~{Zhu} and J.~{Wang}, ``Chunk-based resource allocation in {OFDMA}
  systems-{Part II}: {Joint} chunk, power and bit allocation,'' \emph{IEEE
  Trans. Commun.}, vol.~60, no.~2, pp. 499--509, Feb. 2012.

\bibitem{dahrouj2010coordinated}
H.~Dahrouj and W.~Yu, ``Coordinated beamforming for the multicell multi-antenna
  wireless system,'' \emph{IEEE Trans. Wireless Commun.}, vol.~9, no.~5, pp.
  1748--1759, May 2010.

\bibitem{glorot2010understanding}
X.~Glorot and Y.~Bengio, ``Understanding the difficulty of training deep
  feedforward neural networks,'' in \emph{Proc. int. conf. artificial
  intelligence statistics (AISTATS)}, Sardinia, Italy, May 2010, pp. 249--256.

\bibitem{ba2015adam}
J.~Ba and D.~Kingma, ``Adam: {A} method for stochastic optimization,'' in
  \emph{Proc. Int. Conf. Learning Representations (ICLR)}, San Diego, USA, May
  2015, pp. 1--15.
\end{thebibliography}

\end{document}